\documentclass[aps,prb,twocolumn,amsmath,amssymb,superscriptaddress]{revtex4}

\usepackage{amssymb}
\usepackage{amsmath}
\usepackage{graphicx}
\usepackage{textcomp}
\usepackage{color}
\usepackage{comment}
\usepackage{amsfonts}
\usepackage{epsfig}
\usepackage{hyperref}
\usepackage{bm}
\usepackage[caption=false]{subfig} 
\usepackage{booktabs}
\usepackage{tabularx}
\usepackage{array}
\usepackage{soul}
\usepackage{wasysym}
\usepackage{feyn}
\usepackage{esint}
\usepackage{footmisc}

\newcommand{\arctanh}[1]{\operatorname{arctan}}

\newcolumntype{M}[1]{>{\centering\arraybackslash}m{#1}}

\DeclareMathAlphabet\mathbfcal{OMS}{cmsy}{b}{n}

\begin{document}

\title{The DFT+$\Sigma_2$ method for electron correlation effects at transition metal surfaces}

\author{Andrea Droghetti\footnotemark[3]}
\email[]{andrea.droghetti@tcd.ie}
\affiliation{School of Physics and CRANN, Trinity College, Dublin 2, Ireland}
\author{Milo\v s M. Radonji\'c\footnote[3]{These authors contributed equally to this work.}}
\affiliation{Institute of Physics Belgrade, University of Belgrade, Pregrevica 118, 11080 Belgrade, Serbia}
\author{Anita Halder}
\affiliation{School of Physics and CRANN, Trinity College, Dublin 2, Ireland}
\author{Ivan Rungger}
\affiliation{National Physical Laboratory, Hampton Road, Teddington TW11 0LW, United Kingdom}
\author{Liviu Chioncel}
\affiliation{Theoretical Physics III, Center for Electronic Correlations and Magnetism,
Institute of Physics, and
Augsburg Center for Innovative Technologies, University of Augsburg, 86135 Augsburg, Germany}

\begin{abstract}

We present a computational approach for electronically correlated metallic surfaces and interfaces, which combines Density Functional and Dynamical Mean Field Theory using a multi-orbital perturbative solver for the many-body problem. Our implementation is designed to describe ferromagnetic metallic thin films on a substrate. 
The performances are assessed in detail for a Fe monolayer on a W(110) substrate, a prototypical nanoscale magnetic system. Comparing our results to photoemission data, 
we find qualitative and quantitative improvements in the calculated spectral function with respect to the results of density functional theory within the local spin density approximation. 
In particular, the spin-splitting of the $d$ states is drastically reduced and, at the same time, their spectral width becomes narrower. The method is therefore able to account for the main correlation effects in the system.
\end{abstract}

\maketitle

\section{Introduction}

Metallic heterostructures, formed by stacking different thin films, are the building blocks of spintronic devices, such as spin-valves. Their functionalities are largely determined by the interfaces between the various layers and by the presence of ferromagnetic transition metals (TMs), such as Fe, Co and Ni. Owing to their partially filled $3d$ shells, the electronic structure of these compounds is characterized by electron correlations, which are modified and possibly enhanced at surfaces and interfaces because of atomic relaxation and, moreover, of the reduced atomic coordination. These effects can now be studied in very great detail thanks to the dramatic advancement of spectroscopic~\cite{tu.el.18} and scanning probe techniques~\cite{wi.09}. Yet a proper understanding and interpretation of the data requires accurate {\it ab-initio} simulations.\\
Density Functional Theory (DFT)~\cite{jo.gu.89,kohn.99,jone.15} with its various formulations for the exchange-correlation functionals such as the local spin density approximation (LSDA)~\cite{ba.he.72,vo.wi.80}
or the generalized gradient approximation (GGA)\cite{pe.ch.92,pe.ch.93,pe.bu.96} provides a reasonable description for many of the ground state properties of bulk $3d$ transition metals, but turns out insufficient for describing the excitation spectra of Fe, Ni and Co as measured in photoemission spectroscopy\cite{mo.ma.02,br.mi.06}. 
The DFT band structure drastically overestimates the spin splitting of the $3d$ bands and gives too wide majority spin bands. Furthermore, DFT does not capture 
intrinsic many-electron spectroscopic features, such as satellites~\cite{gu.ba.77}.\\
Electronic correlations can be described within the Hubbard model, which has been extended and combined with DFT to incorporate the realistic description of materials~\cite{im.fu.98}. In the so-called DFT+U method \cite{an.za.91,li.an.95,du.bo.98,co.gi.05}, an effective Hubbard-like interaction is added to the LSDA/GGA exchange-correlation density functional and treated at the static mean-field level.
The method has found widespread use for computational materials design. However, when applied to Fe, Ni and Co, it gives a rigid downward shift of the majority spin bands, while the minority states are maintained at the same positions. 
As a result, the spin-splitting of the bands is even more overestimated than in LSDA/GGA calculations \cite{co.gi.05}, while majority spin $3d$ bands remain too wide. 
Hence, the $U$ static potential actually worsen, instead of improving, the accuracy of DFT for these systems.\\
During the last decade, much progress in the theoretical understanding of $3d$ TMs beyond the limitations of the static mean-field DFT+U picture have been achieved through dynamical mean-field theory (DMFT)~\cite{me.vo.89,ge.ko.96,ko.vo.04,ko.sa.06}. In the so-called LSDA+DMFT scheme~\cite{ko.sa.06,held.07}, LSDA calculations provide the {\it ab-initio} material dependent inputs (orbitals and hopping parameters), while DMFT solves the many-body problem for the local interactions. \\
LSDA+DMFT has been applied to address spectral properties of $3d$ ferromagnetic TMs bulk~\cite{li.ka.01,gr.ma.07} and surfaces\cite{gr.ma.07}, TM alloys~\cite{os.vi.18} and TM compounds~\cite{ka.ir.08}, and to estimate magnetic moments above and below the Curie temperature~\cite{li.ka.01}. Electronic correlation effects have been also investigated in digital magnetic heterostructures~\cite{be.ho.11,ch.le.11} or interfaces containing half-metallic ferromagnets~\cite{ma.he.18,ke.ma.20}. Tunable interfacial properties, which emerge in the presence of electronic correlation, have been found in ferromagnetic heterostructures~\cite{ch.ka.05}.
Last but not least, surface properties have been studied~\cite{gr.ma.07,ma.he.18,ke.ma.20} in connection to linear-response spin-dependent charge transport~\cite{ch.mo.15,mo.ap.17}.\\
Despite all these successes, in practice LSDA+DMFT remains a quite complex and a computationally demanding approach especially for inhomogeneous systems like heterostructures and interfaces. Furthermore, the DMFT solvers~\cite{li.ka.98,ka.li.99,ka.li.02,po.ka.06} typically used for ferromagnetic metals are formulated on the imaginary frequency axis and spectral functions are obtained using numerical analytic continuation schemes to the real frequency axis~\cite{ja.gu.96,sa.98,be.go.00}. The resulting spectra are noisy and sharp features are absent or smeared out. This leads to difficulties when comparing theoretical results to spectroscopic data and when computing charge and spin transport properties via generalized Landauer approaches~\cite{Ja.15,dr.ru.17}, which require the integration of the frequency-dependent transmission function~\cite{ch.mo.15,book1}.  
Hence, DMFT solvers that capture the essential correlation features at a lower computational cost and that are easily implemented using real instead of imaginary frequencies are very valuable. \\
In this paper, we present one of such methods. In particular, we show that the many-electron physics in $3d$ TMs can be treated perturbatively up to the second order in the local Coulomb interaction parameter $U$, while keeping the full manifold of all correlated orbitals provided within DFT.
We name this method DFT+$\Sigma_2$ as second order contributions are calculated diagrammatically on the real frequency axis and accounted for via a self-energy $\Sigma_2$.
Differently from the common implementations of DFT and LSDA+DMFT, which treat bulk unit cells of materials, our implementation is designed for thin films, interfaces and metallic heterostructures in spintronic devices. The typical systems we can describe consist of few correlated atomic layers deposited on a non-correlated semi-infinite substrate.\\
The results of DFT+$\Sigma_2$ are expected to be accurate for moderately correlated systems, such as $3d$ TMs, where $U$ is smaller or comparable to the band width. 
According to the Fermi-liquid theory the net results of a $\Sigma_2$ self-energy is that the spin-splitting of the correlated states is drastically reduced and, at the same time, their spectral width becomes narrower compared to LSDA calculations. 
As a specific example, we study a Fe monolayer on a W(110) substrate, a prototypical ultra-thin magnetic film studied experimentally, and which is used, for example, in tunneling magnetoresistance measurements \cite{ku.bo.02,sa.ar.13}. We find a drastic improvement of the spectral function of the $3d$ Fe orbitals compared to the results of DFT and DFT+U. 
The paper is organized as follows. We present the method and the details of our numerical implementation in Sections II and III. A complete derivation of the $\Sigma_2$ self-energy is presented in the Appendix. In the first part of section IV, we show the results for Fe on W(110) comparing the density of states calculated with DFT, DFT+U and DFT+$\Sigma_2$. In the second part, we assess the results against photomemission spectroscopy data. Finally we conclude highlighting the strengths of the method and anticipating potential future applications. 

\section{Method and implementation}\label{Sec.Method}
We employ the electronic structure code Smeagol~\cite{ro.ga.06,book1}, which implements DFT through the Green's function formalism.
Smeagol uses a linear combination of atomic orbitals (LCAO) basis set $\{\phi_\alpha\}$
and obtains the LSDA Kohn-Sham (KS) Hamiltonian from the DFT package Siesta~\cite{so.ar.02}.
We note however that the equations presented in the following are general and can be readily implemented in
any other DFT code based on the LCAO approach. Each basis orbital in Smeagol and Siesta is characterized by its integer index $\alpha$, which is a collective label including the atom, the orbital, and the angular momentum indices. The orbital index can run over different radial functions corresponding to the same angular momentum following to a multiple-zetas scheme \cite{Szabo,so.ar.02}. The spin-dependent DFT KS Hamiltonian of a system $\hat H^\sigma$ can be represented in its matrix form $H^\sigma$ of elements $H^\sigma_{\alpha\beta}=\langle \phi_\alpha \vert \hat H^\sigma \vert \phi_\beta\rangle$, where  $\sigma=\uparrow, \downarrow$ labels the spin. Since in general the basis orbitals
are non-orthogonal, we also need to introduce the overlap
matrix $S$ of elements $S_{\alpha\beta}=\langle \phi_\alpha\vert \phi_\beta \rangle$ (Ref. \cite{so.ar.02}), which are spin-independent. 

\begin{figure}[h]
\centering\includegraphics[width=0.5\textwidth,clip=true]{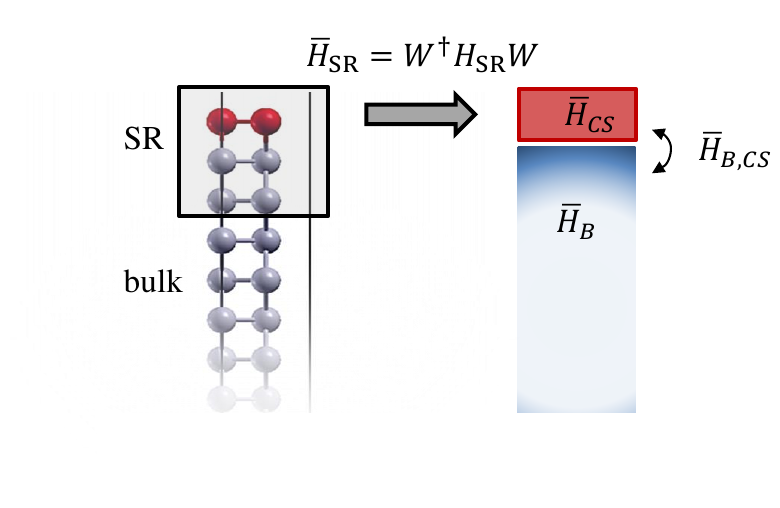}
\caption{One mono-layer of Fe on the W(110).  The red and blue spheres are respectively the Fe and W atoms. 
The surface region is contained in the light blue rectangle. The transformation, Eq. (\ref{eq:Haibath}), which projects the correlated subspace Hamiltonian out of the surface region Hamiltonian is also shown schematically on the right-hand side.}
\label{fig.FeW}
\end{figure}

\subsection{System set-up} 
\label{sec:setup}
The typical system that we study is shown in Fig. \ref{fig.FeW}. It consists of few ferromagnetic TM layers on a semi-infinite non-magnetic substrate. 
The surface is parallel to the $x$-$y$ plane, while its normal vector points along the $z$ direction. 
Periodic boundary condition are applied along $x$ and $y$ so that $\mathbf{k}=(k_x,k_y)$ is the in-plane momentum. 
The system is partitioned into the surface region and the semi-infinite bulk region using a standard approach based on the Green's function technique (e.g. Refs. \cite{ro.ga.06,Ba.Mo.02,ev.we.04,pa.pe.02,book1}). The surface region comprises not only the ferromagnetic layers, but also those few layers of the substrate, which are at the interface and whose electronic structure is different from that of the bulk layers. The surface region is described with $N_\mathrm{SR}$ basis orbitals. \\
The DFT KS Hamiltonian of the system is written as a block matrix\cite{ro.ga.06}
\begin{equation}
\left( \begin{array}{cc}
 H_{\mathrm{SR}}^\sigma(\mathbf{k}) &  H^\sigma_{\mathrm{SR,bulk}}(\mathbf{k})          \\
H^\sigma_{\mathrm{SR,bulk}}(\mathbf{k})^\dagger &   H^\sigma_\mathrm{bulk}(\mathbf{k})          \\
\end{array} \right),\label{h_system}.
\end{equation}
The block $H_{\mathrm{SR}}^\sigma(\mathbf{k})$ is the surface region Hamiltonian, which has dimension $N_\mathrm{SR} \times N_\mathrm{SR}$. $H_{\mathrm{bulk}}(\mathbf{k})$ is the Hamiltonian matrix of the bulk region, whose dimension is infinite because the substrate is semi-infinite along $z$. $H^\sigma_{\mathrm{SR,bulk}}(\mathbf{k})$ is the matrix describing the electronic coupling between the surface and the bulk region.
The Hamiltonian is $\mathbf{k}$-dependent because of the in-plane periodic boundary conditions\cite{so.ar.02}.\\
The orbital overlap matrix of the system has the same structure as the Hamiltonian matrix and reads
\begin{equation}
\left( \begin{array}{cc}
 S_{\mathrm{SR}}(\mathbf{k}) &  S_{\mathrm{SR,bulk}}(\mathbf{k})  \\
S_{\mathrm{SR,bulk}}(\mathbf{k})^\dagger &   S_\mathrm{bulk}(\mathbf{k})          \\
\end{array} \right).\label{S_system}
\end{equation}
$S_\mathrm{SR}(\mathbf{k})$ indicates the $N_\mathrm{SR} \times N_\mathrm{SR}$ overlap matrix of the surface region, $S_{\mathrm{bulk}}(\mathbf{k})$ is the overlap matrix of the bulk region, and $S_{\mathrm{SR,bulk}} (\mathbf{k})$ describes the overlap between the surface and the bulk region.\\
The energy-dependent retarded Green's functions of the surface region is given in the standard form\cite{ro.ga.06, book1}
\begin{equation}
 \mathcal{G}^\sigma_{\mathrm{SR}}(\mathbf{k}; E)= [(E+i\eta)S_\mathrm{SR}(\mathbf{k})-H_\mathrm{SR}^\sigma(\mathbf{k}) -\Delta^\sigma(\mathbf{k};E)]^{-1}\label{GF0}
\end{equation}	
which is a $N_\mathrm{SR} \times N_\mathrm{SR}$ matrix;
$\eta$ is a vanishing small positive real number.
$\Delta^\sigma (\mathbf{k};E)$ is the energy- and momentum-dependent embedding matrix, which describes the hybridization between the SR and the bulk region. It formally reads\cite{ro.ga.06}
\begin{equation}
    \Delta^\sigma(\mathbf{k};E)= K^\sigma(\mathbf{k})^\dagger \mathcal{G}^\sigma_{\mathrm{bulk}}(\mathbf{k};E)K^\sigma(\mathbf{k}),
\end{equation}
where 
\begin{equation}
\mathcal{G}^\sigma_{\mathrm{bulk}}(\mathbf{k};E)=[(E+i\eta)S_\mathrm{bulk}(\mathbf{k})-H_\mathrm{bulk}^\sigma(\mathbf{k})]
\end{equation}
is the retarded Green's function of the bulk region uncoupled from the surface region, and the matrix $K^\sigma(\mathbf{k})$ is
\begin{equation} K^\sigma(\mathbf{k})=[H^\sigma_{\mathrm{SR,bulk}}(\mathbf{k})-(E+i\eta) S_{\mathrm{SR,bulk}}].\end{equation} $\Delta^\sigma(\mathbf{k};E)$ needs to be evaluated via recursive methods (e.g. Refs. \cite{ku.tu.92,na.99}) or semi-analytical methods (e.g. Refs. \cite{ta.gu.01,sa.la.99}). Here we use the algorithm in Ref. \cite{ru.sa.08}.\\
In practice, DFT calculations for the system are performed as follows. First, we obtain the LSDA charge density and KS Hamiltonian of the bulk region, and we compute the embedding matrix $\Delta^\sigma (\mathbf{k};E)$. Then the KS problem for the surface region is solved self-consistently for the boundary conditions set by the bulk\cite{ro.ga.06,book1}.
\subsection{Correlated subspace}\label{Sec.CS}
We assume that there are in total $N_\mathrm{TM}$ atoms in the ferromagnetic layers of the surface region (see Fig. \ref{fig.FeW}). Their $3d$ orbitals span a
$2(5\times N_\mathrm{TM})$-dimensional subspace of the surface region and we call it the ``correlated subspace'' (CS) (the factor $2$ accounts for the spin).
The CS can be projected out from the rest of the system, which we refer to as
the “bath” (B)\cite{ja.ha.09,ja.ha.10,ja.ko.10,Ja.15,dr.ru.17,ap.dr.18} and which includes the orthogonal subspace to the CS within the surface as well as the bulk region. To this aim, we change the basis set.
Specifically, we perform the transformations \cite{dr.ru.17}
  \begin{equation}\begin{split}
\bar{S}_\mathrm{SR}(\mathbf{k})=\left( \begin{array}{cc}
 1 &   0          \\
 0 &   \bar{S}_\mathrm{B} (\mathbf{k})          \\
\end{array} \right)=\\
=W(\mathbf{k})^\dagger S_\mathrm{SR}(\mathbf{k}) W(\mathbf{k}) \end{split}
\label{eq:Saibath}
\end{equation}  
for the SR overlap matrix, 
    \begin{equation}\begin{split}
\bar{H}^\sigma_\mathrm{SR}(\mathbf{k})=\left( \begin{array}{cc}
 \bar{H}^\sigma_\mathrm{CS}(\mathbf{k}) &   \bar{H}^\sigma_\mathrm{CS,B} (\mathbf{k})          \\
 \bar{H}^\sigma_\mathrm{B,CS}(\mathbf{k}) &   \bar{H}^\sigma_\mathrm{B} (\mathbf{k})          \\
\end{array} \right)=\\
=W(\mathbf{k})^\dagger H^\sigma_\mathrm{SR} (\mathbf{k}) W(\mathbf{k}) \end{split}
\label{eq:Haibath}
\end{equation}
for the SR Hamiltonian, and
\begin{equation}\begin{split}
\bar{\mathcal{G}}_\mathrm{SR}^\sigma(\mathbf{k};E)=\left( \begin{array}{cc}
 \bar{\mathcal{G}}^\sigma_\mathrm{CS}(\mathbf{k};E) &   \bar{\mathcal{G}}^\sigma_{\mathrm{CS,B}} (\mathbf{k};E)          \\
 \bar{\mathcal{G}}^\sigma_\mathrm{B,CS}(\mathbf{k};E) &   \bar{\mathcal{G}}^\sigma_\mathrm{B} (\mathbf{k};E)          \\
\end{array} \right)=\\
=W(\mathbf{k})^{-1} \mathcal{G}^\sigma_\mathrm{SR} (\mathbf{k};E) W(\mathbf{k})^{-1^ \dagger},\end{split}
\label{eq:aibath}
\end{equation}
for the SR Green's function. The matrices $W(\mathbf{k})$ are defined in Eq. (10) of  Ref. \cite{dr.ru.17}. 
In the transformed $\bar{S}_\mathrm{SR}(\mathbf{k})$, $\bar{H}_\mathrm{SR}^\sigma(\mathbf{k})$ and $\bar{\mathcal{G}}_\mathrm{SR}^\sigma(\mathbf{k};E)$, 
the top left block describes the CS, the bottom right
block describes the part of the bath included in the surface, and the off-diagonal blocks describe the connection terms. 
We note from Eq. (\ref{eq:Saibath}) that the orbitals of the CS in the transformed basis set become orthogonal and they have zero overlap with the bath orbitals.
$\bar{H}^\sigma_\mathrm{CS}(\mathbf{k})$ in Eq. (\ref{eq:Haibath}) and $\bar{\mathcal{G}}^\sigma_\mathrm{CS}(\mathbf{k};E)$ in Eq. (\ref{eq:aibath})
are respectively the non-interacting Hamiltonian and the non-interacting, or bare, retarded Green's function of the CS. They are matrices of dimension $5N_\mathrm{TM}\times5N_\mathrm{TM}$.\\
In the second quantization formalism the non-interacting CS Hamiltonian reads
\begin{equation}
 \hat{\bar{H}}^\sigma_\mathrm{CS}(\mathbf{k})=\sum_{i,j,\lambda_1, \lambda_2, \sigma}[\bar{ H}^\sigma_{\mathrm{CS}}(\mathbf{k})]_{i \lambda_1, j \lambda_2} \hat d_{i \lambda_1 \sigma}^{\dag} \hat d_{j \lambda_2 \sigma}
\end{equation}
where $\hat d_{i \lambda \sigma}^{\dag}$ and 
$\hat d_{i \lambda \sigma}$ are the electron creation and annihilation operators at orbital $\lambda$ within the atom $i$ and spin $\sigma$ ($i=1,...,N_\mathrm{TM}$ and $\lambda=1,...,5$, $\sigma=\uparrow, \downarrow$).
$[\bar{ H}^\sigma_{\mathrm{CS}}(\mathbf{k})]_{i \lambda_1, j \lambda_2}$ is the CS Hamiltonian matrix element 
between the $d$ orbital $\lambda_1$ of the atom $i$ and the $d$ orbital $\lambda_2$ of the atom $j$.\\
Next, we assume that only electrons in the CS are interacting. 
To describe the Coulomb interaction within the CS, we add the Hubbard-like term to $\hat{\bar{H}}^\sigma_\mathrm{CS}$ as follow
\begin{equation}\begin{split}
& \hat{\bar{H}}^\sigma(\mathbf{k})_{\mathrm{CS},U}= \hat{\bar{H}}^\sigma_\mathrm{CS}(\mathbf{k})+ \\ &+\frac{1}{2}\sum_{\substack{i,\lambda_1,\lambda_2,\lambda_3,\\ \lambda_4,\sigma_1,\sigma_2}} 
U_{\lambda_1,\lambda_2,\lambda_3,\lambda_4} d_{i \lambda_1 \sigma_1}^{\dag} d_{i \lambda_2 \sigma_2}^{\dag} d_{i \lambda_4 \sigma_2} d_{i \lambda_3 \sigma_1}
 -\hat{H}^\sigma_{\mathrm{CS},dc} \end{split}\label{Hint}
\end{equation}
where $U_{\lambda_1,\lambda_2,\lambda_3,\lambda_4}$ are the four-index $U$ parameters, i.e.
the matrix elements of the screened Coulomb interaction between four $3d$ orbitals located at the same site. 
They are parametrized in terms of the average effective Coulomb interaction $U$ and exchange $J$ (Ref. \cite{pavarini})
\begin{eqnarray}
&U=\frac{1}{(2l+1)^2
}\sum_{\lambda_1,\lambda_2}U_{\lambda_1,\lambda_2,\lambda_1,\lambda_2}\label{U}\\
&J=\frac{1}{2l(2l+1) }\sum_{\lambda_1\neq
\lambda_2,\lambda_2}U_{\lambda_1,\lambda_2,\lambda_2,\lambda_1}.\label{J}
\end{eqnarray}
$\hat{H}^\sigma_{\mathrm{CS},dc}$ is the double-counting correction, which is needed to cancel the Coulomb interaction 
already included in the LSDA exchange-correlation potential. 
The form of the double-counting correction is not exactly known, and several approximations have been proposed and used in practice (see for example Ref. \cite{li.ka.01,ko.sa.06,ka.ul.10,ha.ye.10}). 
We will return to this in Sec. \ref{sec.Sigma} after Eq. (\ref{eqLDAU}). \\
The solution of the interacting Hamiltonian $\hat{\bar{H}}^\sigma(\mathbf{k})_{\mathrm{CS},U}$ gives the interacting, or dressed, Green's function $\bar{G}^\sigma_{\mathrm{CS}}(\mathbf{k}, E)$ and the retarded self-energy $\bar{\Sigma}^\sigma_{\mathrm{CS}}(\mathbf{k};E)$ of the CS, which satisfy the Dyson equation
\begin{equation}
\bar{G}^\sigma_{\mathrm{CS}}(\mathbf{k};E)=[\bar{\mathcal{G}}^\sigma_\mathrm{CS}(\mathbf{k};E)^{-1}-\bar{\Sigma}^\sigma_{\mathrm{CS}}(\mathbf{k};E)]^{-1},\label{dressed_G_bar}
\end{equation}
where $\bar{\mathcal{G}}^\sigma_{\mathrm{CS}}(\mathbf{k};E)$ is the bare Green's function of Eq. (\ref{eq:aibath}). $\bar{\Sigma}^\sigma_{\mathrm{CS}}(\mathbf{k};E)$ is evaluated as explained in the following sections.\\
Since the bath is assumed to be non-interacting, the full surface region self-energy in the transformed basis set is
 \begin{equation}
\bar{\Sigma}^\sigma_\mathrm{SR}(\mathbf{k};E)=
\left( \begin{array}{cc}
 \bar{\Sigma}^\sigma_{\mathrm{CS}}(\mathbf{k};E) &   0    \\
 0 & 0         \\
\end{array} \right).\label{SigmaSR}
\end{equation}
It can alternatively be expressed in the original basis by performing the inverse of the transformation in Eq. (\ref{eq:Haibath}), namely\cite{dr.ru.17}
 \begin{equation}
\Sigma^\sigma_\mathrm{SR}(\mathbf{k};E)=W(\mathbf{k})^{-1^\dagger}
\bar{\Sigma}^\sigma_\mathrm{SR}(\mathbf{k};E)
W(\mathbf{k})^{-1 }.\label{SigmaTrans}
\end{equation}
Finally, the dressed Green's function of the surface region in the original basis set is
\begin{equation}
G_\mathrm{SR}^\sigma(\mathbf{k};E)=[\mathcal{G}_\mathrm{SR}^\sigma(\mathbf{k};E)^{-1}-\Sigma_\mathrm{SR}^\sigma(\mathbf{k};E)]^{-1}\label{dressed_G},
\end{equation}
where $\mathcal{G}_\mathrm{SR}^\sigma(\mathbf{k}, E)$ is given in Eq. (\ref{GF0}). From the Green's function we then obtain the density of states (DOS)  
\begin{equation}
\mathrm{DOS}^\sigma(E)=-\frac{1}{N_\mathbf{k}}\sum_{\mathbf{k}}\bigg[\frac{1}{\pi}\mathrm{Im}G_\mathrm{SR}^\sigma(\mathbf{k};E)\bigg],
\end{equation}
where $N_\mathbf{k}$ is the number of $\mathbf{k}$-points.\\
It is important to note that our implementation allows to take into account the effect of the electron-electron interaction as well as that of the electronic coupling between the surface and the non-correlated continuum bulk states. 
The states of the surface region are shifted and broaden owing to the real and imaginary parts of both the self-energy $\Sigma^\sigma_{\mathrm{SR}}(\mathbf{k};E)$ and hybridization function $\Delta^\sigma (\mathbf{k};E)$. The interaction and the hybridization contributions are generally equally important in surface science problems. 

\subsection{DMFT approximation}\label{sec.DMFT}
The self-energy of the CS, $\bar{\Sigma}^\sigma_{\mathrm{CS}}(\mathbf{k};E)$, is a $5N_\mathrm{TM}\times5N_\mathrm{TM}$ matrix, which is energy- and momentum-dependent like the retarded CS Green's function. The calculations accounting for both the $\mathbf{k}$- and the $E$-dependence is a computationally too demanding task for realistic systems. Therefore, we employ the DMFT approximation to simplify the problem, and we consider electron correlation local in space\cite{ko.sa.06,held.07}. 
We assume the self-energy matrix to be momentum independent and to have block-diagonal form 
\begin{equation}
 \bar{\Sigma}^\sigma_{\mathrm{CS},DMFT}(E)=
 \left( \begin{array}{cccc}
 \bar{\Sigma}^\sigma_1(E) &   0 & ... & 0          \\
 0 & \bar{\Sigma}^\sigma_2(E) & ... & 0  
 \\
 ...\\
 0 & 0 & ... & \bar{\Sigma}^\sigma_{N_\mathrm{TM}}(E)            \\
\end{array} \right),
\label{Sigma_local}
\end{equation}
where $\bar{\Sigma}^\sigma_i(E)$ is the $5 \times 5$ block for the $3d$ orbitals of the TM atom $i$. $\bar{\Sigma}^\sigma_i(E)$ may in general be non-diagonal.
The self-energy is evaluated via the self-consistent DMFT procedure\cite{ko.sa.06,held.07}. Our implementation is similar to that suggested by Valli {\it et al.} for model systems~\cite{va.sa.12} and Jacob {\it et al.} for nano-contacts~\cite{ja.ha.10}. The main steps are the following:
\begin{itemize}
\item[i)] We compute the dressed Green's function $\bar{G}^\sigma_{\mathrm{CS}}(\mathbf{k}, E)$ in Eq. (\ref{dressed_G_bar}) with  $\bar{\Sigma}^\sigma_{\mathrm{CS},DMFT}(E)$ instead of $\bar{\Sigma}^\sigma_{\mathrm{CS}}(\mathbf{k};E)$. In the first iteration of the self-consistent procedure we need an initial guess for the $\bar{\Sigma}^\sigma_{\mathrm{CS},DMFT}(E)$. In our calculation we set it to zero.
\item[ii)] We define the so-called local Green's function 
\begin{equation}
  \bar{G}^\sigma_{\mathrm{loc}}(E)= \frac{1}{N_\mathbf{k}}\sum_\mathbf{k} \bar{G}^\sigma_{\mathrm{CS}}(\mathbf{k};E).\label{LocalGreen}
\end{equation}
\item[iii)] We build the dynamical field  $\mathcal{G}^\sigma_{\mathrm{DF},i}(E)$ for each TM atom $i$ inside the CS
\begin{equation}
\mathcal{G}^\sigma_{\mathrm{DF},i}( E)=\{[\bar{G}^\sigma_{\mathrm{loc},i}(E)]^{-1}+ \bar{\Sigma}^\sigma_i(E)\}^{-1}.\label{WeissField}
\end{equation}
where $\bar{G}^\sigma_{\mathrm{loc},i}(E)$ is the $5\times 5$ block of the local Green's function matrix relative to the atom $i$. 
\item[iv)] We map each of the $N_\mathrm{TM}$ atom inside the CS into
an impurity model by defining the bare impurity Green's function of each TM atom as $g^\sigma_{\mathrm{imp},i}(E)=\mathcal{G}^\sigma_{\mathrm{DF},i}(E)$.
\item[v)] We solve the impurity problems as described in the next section
and we get the impurity self-energies $\Sigma^\sigma_{\mathrm{imp},i}(E)$ for each atom $i$. 
\item[vi)] We set $\bar{\Sigma}^\sigma_{i}(E)=\Sigma^\sigma_{\mathrm{imp},i}(E)$ for each atom $i$
and recompute the CS DMFT self-energy in Eq. (\ref{Sigma_local}).
\end{itemize}
These steps are iterated to convergence.\\
Once the self-energy $\bar{\Sigma}^\sigma_{\mathrm{CS},DMFT}(E)$ is computed, the self-energy matrix of the whole surface region is obtained and transformed into the original basis set $\Sigma^\sigma_\mathrm{SR}(\mathbf{k};E)$. We note that in spite of the DMFT approximation used to compute the CS self-energy, the self-energy of the surface region in the original basis $\Sigma^\sigma_\mathrm{SR}(\mathbf{k};E)$ acquires a $\mathbf{k}$-dependence because of the transformation matrices $W(\mathbf{k})$ in Eq. (\ref{SigmaTrans}). 

\subsection{Self-energy in perturbation theory}
\label{sec.Sigma}
We now describe in detail the method used to compute the self-energy by solving the impurity problem. In case of weakly or moderately correlated systems, such as the $3d$ ferromagnetic metals of interest here, the self-energy can be obtained using self-consistent perturbative approaches in terms of skeleton diagrams around the non-interacting solution. A popular scheme is the self-consistent fluctuating exchange approximation (FLEX)~\cite{bi.sc.89}, which is conserving in the Baym-Kadanoff sense~\cite{ba.ka.61,baym.62,bookStefanucci}.
The FLEX has been further combined with the $T$-matrix approximation, into the spin-polarized $T$-matrix fluctuating exchange approximation~\cite{li.ka.98,ka.li.99,ka.li.02,po.ka.06}, the so-called SPT-FLEX impurity solver. 
The SPT-FLEX self-energy is described by the Hartree and Fock diagrams with the formal replacement of the bare interaction by the $T$-matrix \cite{gali.58,ka.63}, which obeys the Bethe-Salpether-like integral equation. 
The SPT-FLEX is formulated on the imaginary (Matsubara) frequency axis and correlation functions are obtained indirectly via the numerical analytical continuation to the real frequency axis.
Unfortunately, this often leads to numerical difficulties since the analytical continuation of discrete numerical data is not unambiguous and, in addition, it requires the treatment of the high-frequency ``tails''~\cite{se.he.91}. For this reason, we consider here a simpler second order perturbative treatment implemented to provide the self-energy directly on the real frequency axis, while retaining the multi-orbital nature of the many-body problem. A similar approach was used by Drchal {\it et al.} to study TM alloys\cite{dr.ja.99}. The method represents a good compromise between the need for an accurate description of correlated effects and for a efficient and easy numerical evaluation of spectral properties.  \\
We assume that the matrix $g^\sigma_{\mathrm{imp},i}(E)$ for the impurity problem associated to the TM atom $i$ in Eq. (\ref{WeissField}) is diagonal. This greatly reduces the computational effort of the calculations. We denote the diagonal elements of the impurity bare Green's function and self-energy as $g^\sigma_{i\lambda}(E)\equiv[g^\sigma_{\mathrm{imp},i}(E)]_{\lambda,\lambda}$ and $\Sigma^\sigma_{i\lambda}(E)\equiv[\Sigma^\sigma_{\mathrm{imp},i}(E)]_{\lambda,\lambda}$, where $\lambda=1, ...,5$ labels the $3d$ orbitals. The Dyson equation for the impurity problem is then 
\begin{equation}\label{eq.Dyson_impurity}
    G^\sigma_{i\lambda}(E)^{-1}=g^\sigma_{i\lambda}(E)^{-1}-\Sigma^\sigma_{i\lambda}(E)
\end{equation}
where $G^\sigma_{i\lambda}(E)^{-1}$ is the dressed Green's function for orbital $\lambda$ of the the impurity $i$.
The self-energy up to the second order in perturbation theory in $U$ over the band width is obtained by using the skeleton diagrams in appendix \ref{app.diagrams},
 and is written as
\begin{equation}\label{eq:sigma_1+2}
 \Sigma^\sigma_{i \lambda }(E)\approx \Sigma^{\sigma (1)}_{i \lambda }+\Sigma^{\sigma (2)}_{i \lambda }(E).
\end{equation}
The first-order contribution 
\begin{equation}
\Sigma^{\sigma (1) }_{i \lambda } = 
\sum_{\lambda_1 \sigma_1} U_{\lambda\lambda_1\lambda\lambda_1} n_{i \lambda_1}^{\sigma_1} - \sum_{\lambda_1} U_{\lambda\lambda_1\lambda_1\lambda} n_{i \lambda_1}^{\sigma}.\label{eq.first}
\end{equation}
is the well-known Hartree-Fock approximation, where 
\begin{equation}
  n_{i \lambda}^{\sigma}=\int_{-\infty}^\infty dE f(\omega) \textrm{Im}G^{\sigma}_{i \lambda }(E)\label{eq.occupation}  
\end{equation}
is the occupation of the orbital $\lambda$ of spin $\sigma$ at the atom $i$; $f(E)$ is the Fermi function.
$\Sigma^{\sigma (1)}_{i \lambda }$ is local in time, i.e. energy independent. It therefore represents a one-electron potential producing only a shift of the non-interacting energy levels.\\
The second order contribution can be split into its real and imaginary part. The imaginary part is given by
\begin{widetext}
\begin{eqnarray}
\label{eq:sig2r}
\mathrm{Im}\left[\Sigma^{\sigma (2)}_{i \lambda}(E)\right] &= 
-\pi \sum\limits_{\lambda_1\lambda_2\lambda_3\sigma_1}& U_{\lambda\lambda_1\lambda_2\lambda_3}U_{\lambda_3\lambda_2\lambda_1\lambda} 
\int_{-\infty}^\infty \mathrm{d}\epsilon_1 \int_{-\infty}^\infty\mathrm{d} \epsilon_2 
D_{i\lambda_1}^{\sigma_1}(\epsilon_1) 
D_{i\lambda_2}^{\sigma}(\epsilon_2) 
D_{i\lambda_3}^{\sigma_1}(\epsilon_1+\epsilon_2-E) \times \nonumber \\
&&\{ f(\epsilon_1) f(\epsilon_2) + 
\left[ 1-f(\epsilon_1) - f(\epsilon_2) \right] f(\epsilon_1+\epsilon_2-E) \} \nonumber \\
&+\pi \sum\limits_{\lambda_1\lambda_2\lambda_3}& U_{\lambda\lambda_1\lambda_2\lambda_3}U_{\lambda_2\lambda_3\lambda_1\lambda}
\int_{-\infty}^\infty \mathrm{d}\epsilon_1 \int_{-\infty}^\infty\mathrm{d} \epsilon_2 
D_{i\lambda_1}^{\sigma}(\epsilon_1+\epsilon_2-E) 
D_{i\lambda_2}^{\sigma}(\epsilon_2) 
D_{i\lambda_3}^{\sigma}(\epsilon_1) \times \nonumber  \\
&&\{ f(\epsilon_2) f(\epsilon_1) + 
\left[ 1-f(\epsilon_2) - f(\epsilon_1) \right] f(\epsilon_1+\epsilon_2-E) \},\label{eq.second}
\end{eqnarray}
\end{widetext}
 where 
 \begin{equation}
D_{i \lambda}^\sigma(E) = -\frac{1}{\pi} \mathrm{Im}G_{i \lambda}^{\sigma }(E)] \label{eq.D_i}   
\end{equation}
is the spectral function of $G_{i \lambda}^{\sigma }(E)$. 
The real part is given by the Kramers-Kronig relations
\begin{equation}
    \mathrm{Re} \left[ \Sigma^{\sigma (2)}_{i \lambda}(E) \right] = -\frac{1}{\pi} \int_{-\infty}^{\infty} d\epsilon \frac{\mathrm{Im}\left[\Sigma^{\sigma (2)}_{i \lambda}(\epsilon)\right]}{E-\epsilon}.\label{eq.KK}
\end{equation}
Eqs. (\ref{eq.second}) and (\ref{eq.KK}) are easily implemented thus allowing for the calculation of the second order self-energy contributions. \\
The approximation for the self-energy that we introduced is called second Born-approximation \cite{bookStefanucci} and it is conserving in the Baym-Kadanoff sense. Eq. (\ref{eq.Dyson_impurity}) needs to be solved self-consistently together with Eqs. (\ref{eq.first}) and  (\ref{eq.second}) because the self-energy is a functional of the dressed impurity Green's function, that is $\Sigma^{\sigma (1)}_{i \lambda }=\Sigma^{\sigma (1)}_{i \lambda }[G^\sigma_{i\lambda}]$ and $\Sigma^{\sigma (2)}_{i \lambda }=\Sigma^{\sigma (2)}_{i \lambda }[G^\sigma_{i\lambda}]$.\\ 
Eqs. (\ref{eq.second}) and (\ref{eq.KK}) are easily implemented thus allowing for the calculation of the second order self-energy contributions. 
The numerical integration over the frequencies $\epsilon_1$ and $\epsilon_2$ in Eq. (\ref{eq.second}) would potentially represent a computational bottleneck of the method if it was carried out using a too large number of discrete frequency points.
However, in the case of ferromagnetic TMs, we find accurate results already for relatively coarse energy grids ($dE\approx10^{-3}$ eV) thus making the evaluation of $\Sigma^{\sigma (2)}_{i \lambda }(E)$ computationally quite inexpensive. \\
\begin{figure*}[ht]
\includegraphics[width=0.9\linewidth]{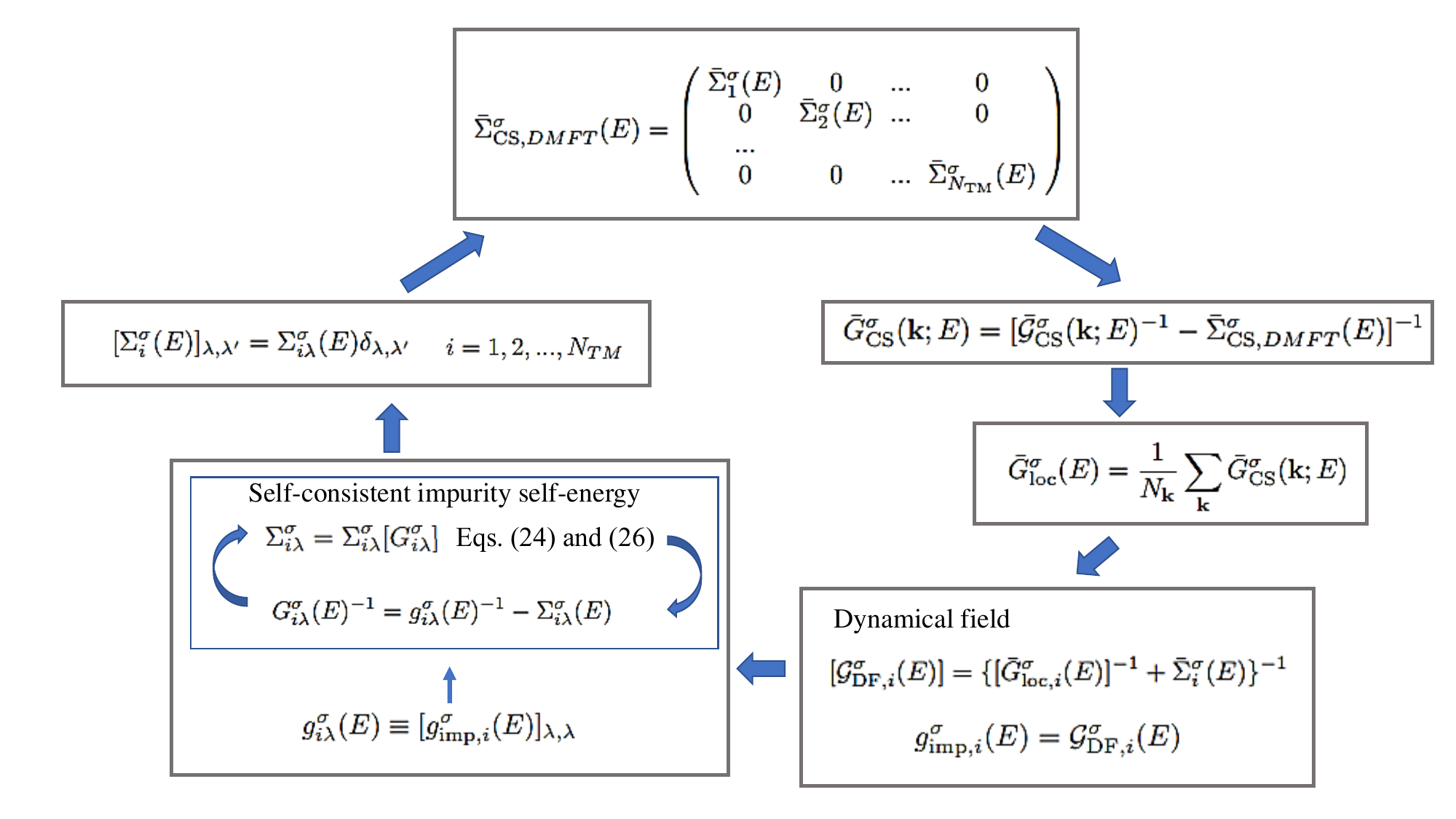}
\caption{Schematic representation of the self-consistent $\Sigma_2$ calculations.}\label{Fig.DMFTloop}
\end{figure*}

\subsection{Self-energy calculations combined with DFT}\label{sec.DFT_Sigma}
We now discuss how self-energy calculations are combined with DFT, and how the first and second order contributions are taken into account in our numerical implementation. Different scheme are proposed to carry out calculations.
\subsubsection{The reduction to LSDA+U}\label{sec.LSDAU}
The first order term of Eq.~(\ref{eq:sigma_1+2}) combined with DFT reduces to the LSDA+U approach. 
The double counting correction $H^\sigma_{\mathrm{CS},dc}$ of Eq. (\ref{Hint}) can be englobed into $\Sigma^{\sigma (1)}_{i \lambda}$ and can be approximated with one of the various forms proposed for LSDA+U \cite{bh.af.14}.
For example, using the so-called fully-localized limit~\cite{cz.sa.94} for $H^\sigma_{\mathrm{CS},dc}$, leads to 
\begin{equation}\begin{split}
 &\Sigma^{\sigma (1)}_{i \lambda ,dc} = \Sigma^{\sigma (1)}_{i \lambda} -H^\sigma_{\mathrm{CS},dc} =
\sum_{\lambda_1 \sigma_1} U_{\lambda\lambda_1\lambda\lambda_1} n_{i \lambda_1}^{\sigma_1} \\
- &\sum_{\lambda_1} U_{\lambda\lambda_1\lambda_1\lambda} n_{i \lambda_1}^{\sigma}
-\Big[U\Big(N_i-\frac{1}{2}\Big)+J\Big(N_i^\sigma-\frac{1}{2}\Big)\Big],\end{split}\label{eqLDAU}
\end{equation}
where $N_i=\sum_\sigma N_i^\sigma=\sum_{\sigma\lambda} n_{i \lambda}^{\sigma}$ is the total occupation, and $U$ and $J$ are
the average effective Coulomb and exchange interactions in Eqs. (\ref{U}) and (\ref{J}).
$\Sigma^{\sigma (1)}_{i \lambda ,dc}$ is exactly the Hubbard corrective potential of the LSDA+U Hamiltonian proposed by Lichtenstein {\it et al.} in Ref.~\cite{li.an.95} 
[note that, differently from the original formulation, only diagonal elements of the density matrix $n_{i \lambda_1}^{\sigma_1}$ appear in Eq. (\ref{eqLDAU}) because we assumed the Green's function to be diagonal in the orbital indices].
Based on these observations, we evaluate the first-order self-energy performing a standard LSDA+U calculation.
Although, Eq. (\ref{eqLDAU}) is the most complete formulation of the LSDA+U Hubbard corrective potential with fully orbital-dependent electronic interactions, we use a simplified expression introduced by Dudarev {\it et al.}~\cite{du.bo.98}  
\begin{equation}
\Sigma^{\sigma (1)}_{i \lambda,dc } \approx V_{U,i\lambda}^\sigma=(U-J)(\frac{1}{2}-n_{i\lambda}^\sigma).\label{eq.Dudarev}
\end{equation}
to reduce the complexity of the calculations. 
This simplified expression has been successfully applied in several studies and for most materials it yields similar results as the fully rotationally invariant formulation (see Ref. \cite{bh.af.14} and references therein for more details).

\subsubsection{The LSDA+$\Sigma_2$ scheme}\label{sec.lsda_sigma}
The self-energy can be evaluated in terms of the bare impurity Green's function instead of the dressed impurity Green's function, i.e. $\Sigma^{\sigma (1)}_{i \lambda }=\Sigma^{\sigma (1)}_{i \lambda }[g^\sigma_{i\lambda}]$ and $\Sigma^{\sigma (2)}_{i \lambda }=\Sigma^{\sigma (2)}_{i \lambda }[g^\sigma_{i\lambda}]$ with $g^\sigma_{i\lambda}$ that
replaces $G^\sigma_{i\lambda}$ in Eqs. (\ref{eq.occupation}) and (\ref{eq.D_i}).
The approach has already been used in the literature \cite{li.li.00}, and in particular in Refs.~\cite{kl.me.81,tr.du82,alig.06}, although these papers consider a single-orbital model~\cite{alig.06} or an average interaction $U$ (Ref.~\cite{tr.du82}), while we maintain a multi-orbital description including the complete four-index interaction. The approximation using bare Green's functions neglects some the diagrams in the perturbative expansion of the self-energy compared to the second Born-approximation\cite{bookStefanucci} and it is non-conserving in the Baym-Kadanof sense. Nonetheless, we find that both approaches provide similar total number of electrons.\\
The fact that the Hartree-Fock self-energy is static leaves us the freedom to chose how to practically perform the perturbative calculations up to second order. Namely, starting from the non-interacting Green's function $g^\sigma_{i \lambda }( E)$ we can calculate the total self-energy corrections up to second order, i.e. including the static $\Sigma^{\sigma (1)}_{i \lambda }$ and the dynamic $\Sigma^{\sigma (2)}_{i \lambda }(E)$ 
contributions. Alternatively, we can "immerse" the static contribution, approximated as in Eq. (\ref{eq.Dudarev}), into the non-interacting local Green's functions. This means that
$g^\sigma_{i \lambda }( E)$ is replaced by the LSDA+U Green's function 
\begin{equation}
 g^\sigma_{\mathrm{LSDA+U},i \lambda }( E)= [g^\sigma_{i \lambda }( \omega)^{-1}-V_{U,i\lambda}^\sigma]^{-1}.\label{GF_LDAU}
\end{equation}
The Dyson Eq. (\ref{eq.Dyson_impurity}) 
retains its structure with the total self-energy $\Sigma^\sigma_{i \lambda}(E)$ substituted by the correlation self-energy $\Sigma^\sigma_{C,i \lambda}(E)=\Sigma^\sigma_{i \lambda }(E)- V_{U,i\lambda}^\sigma$, where 
$\Sigma^\sigma_{C,i \lambda}(E)$ is evaluated using the LSDA+U Green's functions $g^\sigma_{\mathrm{LSDA+U},i \lambda }( E)$. \\
Following these considerations, our calculations are practically carried out in the following way. 
We perform a fully charge self-consistent LSDA+U calculation and we obtain $V_{U,i\lambda}^\sigma$ with $n^\sigma_{i\lambda}$ the LSDA+U occupation of atom $i$ and orbital $\lambda$. Then, in the first DMFT iteration, we use $g^\sigma_{\mathrm{LSDA+U},i \lambda }( E)$ as impurity Green's function, which already contains the Hartree-Fock part of the self-energy, while we compute the second order self-energy corrections inserting $g^\sigma_{\mathrm{LSDA+U},i \lambda }( E)$ into Eq. (\ref{eq.second}). The DMFT equations are eventually iterated solving the impurity problem by re-evaluating $\Sigma^{\sigma (1)}_{i \lambda }$ and $\Sigma^{\sigma (2)}_{i \lambda }(E)$ at each DMFT iteration. For the systems of interest here, and in the limit of weak interactions, we found the DMFT self-consistent procedure converges very fast, and changes in the DOS are negligible after the first few DMFT iteration. In fact perturbative corrections in metallic systems lead to moderate changes of the electronic structure. The system remains a well defined Fermi liquid with modified parameters.
We refer to such computational approach as LSDA+$\Sigma_2$. Calculations carried out by means LSDA+$\Sigma_2$ add just a small computational cost to standard LSDA+U calculations, while already capturing some correlation effects as shown in Sec. \ref{Sec.Results}. 
\subsubsection{Self-consistent $\Sigma_2$ approximation}\label{sec.lsda_SCsigma}
The self-energy terms are evaluated as functionals of the dressed impurity Green's function, that is $\Sigma^{\sigma (1)}_{i \lambda }[G^\sigma_{i\lambda}]$ and $\Sigma^{\sigma (2)}_{i \lambda }[G^\sigma_{i\lambda}]$. A calculation requires in practice two self-consistent cycles as shown in Fig. \ref{Fig.DMFTloop}. The first, is the DMFT cycle explained in Sec. \ref{sec.DMFT}. Then at each DMFT iteration, a second self-consistent cycle is done to solve the impurity problem as explained in Sec. \ref{sec.Sigma}. 
To ensure electron conservation inside the impurity according to the Luttinger-Ward theorem \cite{bookStefanucci}, we add an identical on-site potential $v$ to all correlated $3d$ orbitals and we readjust $v$ at each iteration. 

\section{Computational details}\label{Sec.Details}
\begin{figure}[h]
\centering\includegraphics[width=\columnwidth,clip=true]{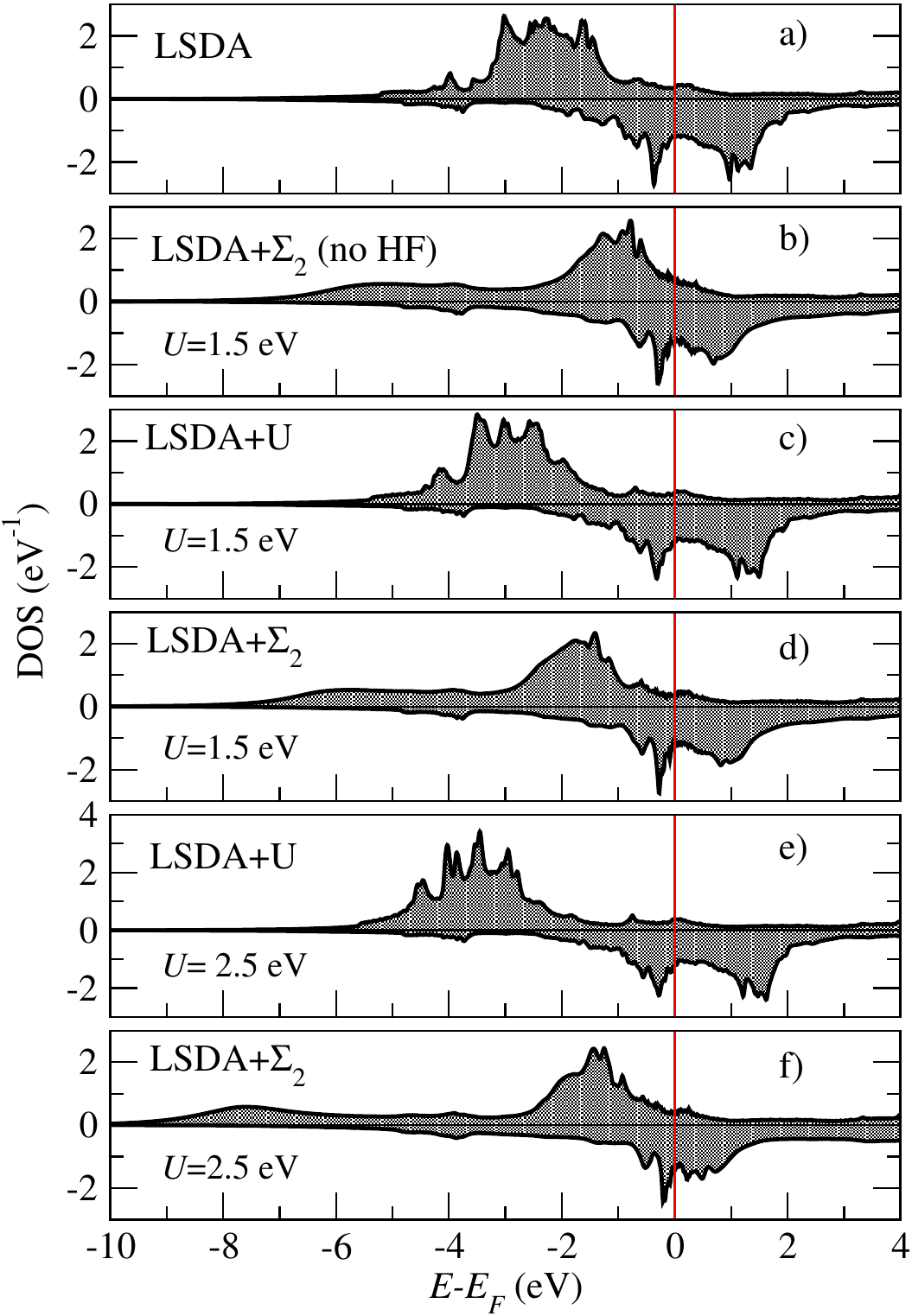}
\caption{DOS of one of the Fe atoms calculated by means of LSDA, LSDA+U and LSDA+$\Sigma_2$. The Fermi level is at 0 eV. The value of $U$ used in LSDA+U is indicated. $J =0.5$ eV in all calculations. LSDA+$\Sigma_2$ (no HF) indicates calculations where the approximate HF potential of Eq. \ref{eq.Dudarev} is neglected. }
\label{fig.DOS}
\end{figure}

We treat core electrons with norm-conserving Troullier-Martin pseudopotentials. 
The valence states are expanded through a numerical atomic orbital basis set including multiple-$\zeta$ and polarized functions~\cite{so.ar.02}. 
The electronic temperature is set to $300$ K.
The real space mesh is set by an equivalent energy cutoff of $300$ Ry. We use $15\times15$ $\mathbf{k}$-point mesh to compute the self-consistent charge density with LSDA(+U). 
This charge density is then used as input in a non-self-consistent calculation
to obtain the density of states employing $61\times61$ $\mathbf{k}$-points. We shift all energies in such a way to set the Fermi level at 0 eV.
To calculate the second-order self-energy, we use a frequency grid comprising 4400 points and extending from $-16$ to $6$ eV. The imaginary part $\eta$ in Eq. (\ref{GF0}) is 0.01 eV and 0.005 eV in LSDA(+U) and in LSDA+$\Sigma_2$
calculations, respectively. This leads to an additional broadening of the DOS for a better display in Figs. \ref{fig.DOS} and \ref{fig.d-resolved-DOS}.\\
We express the Coulomb parameters $U_{\lambda_1,\lambda_2,\lambda_3,\lambda_4}$ in terms of Slater integrals $F^0$, $F^2$ and $F^4$ (Ref.~\onlinecite{im.fu.98}). 
These are connected to the average effective Coulomb and exchange interactions of Eqs. (\ref{U}) and (\ref{J}) through the relations $U=F^0$ and 
$J=(F^2+F^4)/14$. The ratio $F^4/F^2$ is assumed to correspond to the atomic value $\approx 0.625$~\cite{an.gu.91}.\\

\section{Electronic structure of Fe/W(110)}\label{Sec.Results}
To describe the performances of our method we consider a monolayer of Fe on a W(110) substrate, a systems quite investigated in nanoscale magnetism and often used in spin-polarized scanning tunneling experiments \cite{ku.bo.02,sa.ar.13}. The system is presented in Fig. \ref{fig.FeW}. Only the results of spin-polarized calculations in the ferromagnetic phase are described, while non-magnetic calculations are shown in Appendix \ref{app.nonmagnetic}. We remark that our method treats the W substrate as truly semi-infinite. This is an advantage  compared to standard implementations of DFT and DFT+DMFT, where Fe/W(110) would be described as a slab with few W layers and therefore introducing unwanted confinement effects in the system electronic structure. \\   

\subsection{Density of states}\label{sec.DOS}

\begin{figure}[h]
\centering\includegraphics[width=\columnwidth,clip=true]{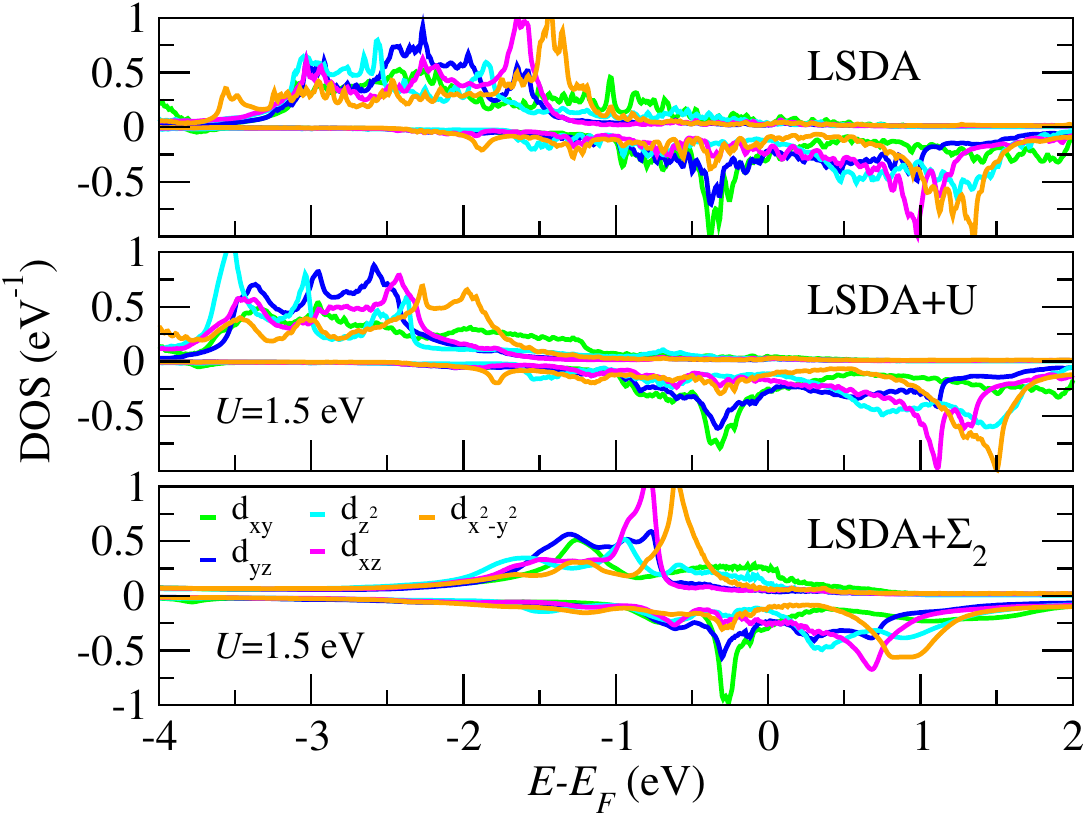}
\caption{Orbital resolved DOS of Fe atom obtained via LSDA and LSDA+$\Sigma_2$ calculations. The Fermi level is at 0 eV.}
\label{fig.d-resolved-DOS}
\end{figure}

 The Fe DOS calculated by means of LSDA, LSDA+U and LSDA+$\Sigma_2$ is presented in Fig. \ref{fig.DOS}. By analysing the results of the different methods we will understand the effect of the dynamical self-energy compared to mean-field static approximations. \\
 To begin with, we observe that already the DOS obtained by using LSDA(+U) has an intrinsic broadening in our calculations. This is due to the embedding matrix function $\Delta^\sigma(\mathbf{k};E)$ between the surface region and the W bulk continuum states in Eq. (\ref{GF0}). The effect would be absent in calculations considering a slab geometry for Fe/W(110), and it demonstrates the importance of using our implementation of DFT to accurately describe surfaces and interfaces. \\
 The LSDA results are shown in Fig. \ref{fig.DOS}-a. The spin up (majority) and spin down (minority) DOS are split by about 2 eV (Fig. \ref{fig.DOS}-a). Comparing with the orbital resolved DOS in Fig.~\ref{fig.d-resolved-DOS}, we find that the prominent peaks for both spin channels have mostly $d$ character. 
The total DOS for the spin up channel is centered at $E-E_F \approx -2.2$ eV and is almost completely filled. The total occupation of the 
majority $d$ orbitals 
is 4.68 electrons.
In contrast, the spin down DOS presents two main peaks at the opposite sides of the Fermi level and separated by a pseudo-gap. 
As seen in the orbital-resolved DOS in Fig.~\ref{fig.d-resolved-DOS}, the two-peak structure reflects the separation of the Fe $d_{xy}$ and $d_{yz}$ from the $d_{xz}$, $d_{z^2}$ and $d_{x^2-y^2}$ orbitals due to the (110) surface symmetry. 
The $d_{xy}$ orbitals are oriented along the crystal direction connecting the surface Fe atoms and thus merge into $\sigma$ bonding and anti-bonding bands.
The Fe $d_{yz}$ orbitals overlap with the $5d_{yz}$ orbitals of the W atoms underneath forming a second set of $\sigma$ bonding and anti-bonding bands.
All bonding states are centered at about $E-E_F\approx -0.4$ eV giving the first sharp peak in the spin down DOS, while the antibonding states are unoccupied and emerge as very broad features extending up to $2$ eV.
The $d_{xz}$, $d_{z^2}$ and $d_{x^2-y^2}$ orbitals mostly overlap with $s$ orbitals forming broad states centered around $E-E_F \approx 1.15$ eV. They contribute to the second unoccupied peak in the spin down DOS.
The occupation of the Fe $d_{xy}$ and $d_{yz}$ orbitals is about $0.46$, whereas that of the $d_{xz}$, $d_{z^2}$ and $d_{x^2-y^2}$ orbitals is $0.31$, contributing to the total occupation of the minority $d$ states equal to  $1.86$ electrons. 
The spin magnetic moment $\mu$ is $2.84$ $\mu_B$, considerably enhanced with respect to that in bulk Fe, 2.2 $\mu_B$. 

The DOS calculated by means of LSDA+U is shown in Fig. \ref{fig.DOS}-c and -e for $U=1.5$ eV and $U=2.5$ eV, respectively, and $J=0.5$ eV.
In such a mean-field like approach the $V_{U,i\lambda}^\sigma$ potential shifts the spin up LSDA DOS towards lower energies by about $-0.5(U-J)$ [see Eq.~(\ref{eq.Dudarev})]. 
The occupation of the spin up states therefore increases as a function of $U$.
In contrast, the $V_{U,i\lambda}^\sigma$ potential barely affects the spin down $d_{xy}$ and $d_{yz}$ orbitals, and moves the $d_{xz}$, $d_{z^2}$ and $d_{x^2-y^2}$ orbitals towards higher energies by about $0.2(U-J)$.
The pseudo-gap in the spin down DOS across the Fermi level widens, while the
$d_{xz}$, $d_{z^2}$ and $d_{x^2-y^2}$ orbitals are slightly emptied. Overall, the splitting between the spin up and down DOS becomes larger 
and, as a result, the total Fe magnetic moment systematically increases as a function of $U-J$. 
This is a general outcome of DFT+U calculations found for all ferromagnetic materials~\cite{co.gi.05}. 
It is a consequence of the fact that the $V_{U,i\lambda}^\sigma$ potential of Eqs.~(\ref{eqLDAU}) and (\ref{eq.Dudarev}) represents the Hartree-Fock approximation to the Hubbard interaction.\\
Next, we discuss the results obtained including second order self-energy contributions via the approach described in Sec. \ref{sec.lsda_sigma}. The DOS are shown in Fig. \ref{fig.DOS}-b, -d and -f. We use $U=1.5$ eV and $U=2.5$ eV and $J=0.5$ eV like in the LSDA+U calculations.
Since the LSDA $d$ bandwidth is about 4.5 eV, the perturbation expansion is valid.
We note that the Fe atoms Green's function is not diagonal. However the off-diagonal elements are typically two orders of magnitude smaller than diagonal elements. They can therefore be discarded, consistent with our assumption in Sec. \ref{sec.Sigma}. 
The calculations in Fig. \ref{fig.DOS}-b, which are indicated as ``LSDA+$\Sigma_2$ (no HF)'', are carried out neglecting the approximate Hartree-Fock potential of Eq. (\ref{eq.Dudarev}) and using the LSDA Green's function to evaluate second order self-energy contributions.\\ 
The main features due to the second order self-energy are similar across Fig.~\ref{fig.DOS}-b, -d and -f. Some redistribution of the spectral weight occurs resulting in a considerable spectral narrowing.
The orbital ordering (orbitals' position in energy) and the character of the main peaks recognizable in Fig.~\ref{fig.d-resolved-DOS} is preserved as it is dictated by the surface symmetry and the crystal field. 
The changes in the DOS are more pronounced for spin up than for spin down. The top of the occupied $d$ DOS
in the spin up channel is shifted towards the Fermi level, while the down DOS is barely affected. 
This leads to a lowering of the Fe magnetic moment $\mu$ reported in Tab.~\ref{Tab.occupation}.
For $U= 2.5$ eV, we observe a reduction in excess of $0.3 \mu_B$ compared to LSDA+U thus
demonstrating that correlation effects play a crucial role in counterbalancing the exchange interaction. This can be easily understood by inspecting the mathematical form of the second order self-energy of Eq.~(\ref{eq.second}) 
or the corresponding Feynman diagrams in the appendix.
Electronic correlations are build-up by the scattering processes between electrons and electron-hole pairs.
Since the spin up channel is essentially full, electron-hole pairs are mostly created in the spin down channel.
Therefore a spin up electron will more likely scatter with an opposite-spin pair costing an effective interaction $U$.
In contrast, a spin down electron will scatter with same-spin pairs with effective interaction $U- J < U$. 
Therefore, correlation effects are stronger for spin up than for spin down electrons. \\
\begin{table}
{
\begin{tabular}{ M{3.5cm} | M{1.2cm}|M{1.2cm}|M{1.2cm} }\hline
  Method   &\space $n_\uparrow$ & $n_\downarrow$ & $\mu(\mu_B)$ \\\hline
 LSDA  &\space  4.68  & \space 1.85 &\space 2.84\\\hline
  LSDA+$\Sigma_2$ (no HF) & \space 4.57 &\space 1.96  &\space 2.6  \\\hline
   LSDA+U ($U=1.5$~eV)  &\space  4.75  & \space 1.75 &\space 3.0 \\\hline
  LSDA+$\Sigma_2$ ($U=1.5$~eV) & \space 4.69 &\space 1.81  &\space  2.88 \\\hline
 LSDA+U ($U=2.5$~eV)  &\space 4.79   & \space 1.69  &\space 3.1 \\\hline
  LSDA+$\Sigma_2$ ($U=2.5$~eV) & \space 4.65 &\space 1.88  &\space  2.76 \\\hline
\end{tabular}}
\caption{Total $d$ orbitals occupation and magnetic moment $\mu=n_\uparrow-n_\downarrow$ (in Bohr magneton $\mu_B$.)}\label{Tab.occupation}
\end{table}
Analyzing in more details the results for $U=1.5$~eV, we note that the spin up $d$ DOS center is predicted at $E-E_F\approx-1.5$~eV and at $\approx-2.0$~eV respectively in calculations without and with the HF potential (Fig.~\ref{fig.DOS}-b and~\ref{fig.DOS}-d). Clearly the differences in the two cases are due to the initial state dependence of perturbation theory and they reflect the differences between the LSDA and LSDA+U DOS. In spite of that, the two results are overall quite similar. 
In this specific case, neglecting the HF potential and using the LSDA Green's function to compute $\Sigma^{\sigma (2)}_{i \lambda }(E)$ is a practical and reliable simplification. \\
Next, w compare the LSDA+$\Sigma_2$ results for $U=1.5$ eV and for $U=2.5$ eV (Fig.~\ref{fig.DOS}-d and -f respectively). We see that a larger $U$ does not significantly change the position of the $d$ states, but it induces a narrowing of both the spin up and the spin down DOS features.
According to this finding, an increase in the local Coulomb parameter $U$ including dynamic self-energy effects leads to a re-shaping of the spectrum rather than in a modification of their spin-splitting as one would expect based on a static mean-field picture. \\
For energies far below from the Fermi level, we distinguish a satellite appearing in the spin-up channel below -4 eV. 
A similar feature has been unequivocally observed in photoemission measurements for Ni~\cite{gu.ba.77} and it has been predicted for bulk Fe as well~\cite{gr.ma.07}. Here we predict that it is present even in a Fe monolayer. 
As already noted in Ref.~\cite{gr.ma.07} for bulk calculations, albeit performed with a SPT-FLEX impurity solver instead of $\Sigma_2$, the center of the satellite systematically shifts towards lower energies when increasing $U$.\\
\begin{figure}[h]
\centering\includegraphics[width=\columnwidth,clip=true]{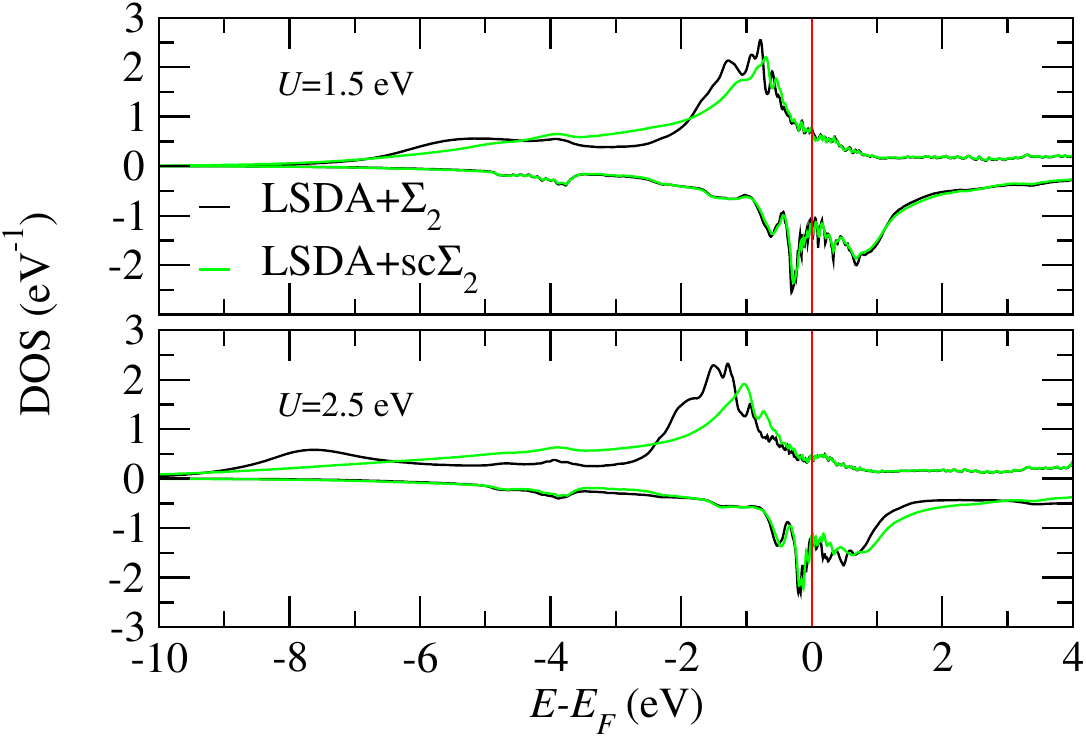}
\caption{Comparison between the DOS calculated LSDA+$\Sigma_2$ and LSDA+sc$\Sigma_2$.}
\label{fig.scSigma2}
\end{figure}
Finally, we present the results of calculations, referred to as LSDA+sc$\Sigma_2$, where the self-energy is evaluated self-consistently as explained in Sec. \ref{sec.lsda_SCsigma}.
The DOS is shown in Fig. \ref{fig.scSigma2} for $U=1.5$ and $U=2.5$ eV ($J=0.5$ eV in both cases). The self-consistent iterations lead to a redistribution of the spin up spectral weight, in particular at energies far from the Fermi level.
The spin down DOS remains almost unaltered owing to the low correlation. For $U=1.5$ eV, the spin-splitting of the $3d$ states remains equal to that predicted in the LSDA+$\Sigma_2$ calculations. In contrast, for $U=2.5$ eV, the spin up $3d$ states are moved towards the Fermi energy by about 0.2 eV compared to non-self-consistent calculations. Thus, the DOS spin-splitting is reduced. Overall, we find that the self-consistent procedure mitigates the initial state dependence of perturbation theory, although it does not completely eliminate it. Notably, at energies far below the Fermi level, the satellite feature becomes much less marked in the LSDA+sc$\Sigma_2$ than in LSDA+$\Sigma_2$ DOS. The self-consistent calculation enhances multi-band screening effects thus leading to a partial suppression of that intrinsic many-body feature. 
\subsection{Self-energy and effective masses}
\begin{figure}[h]
\centering\includegraphics[width=0.47\textwidth,clip=true]{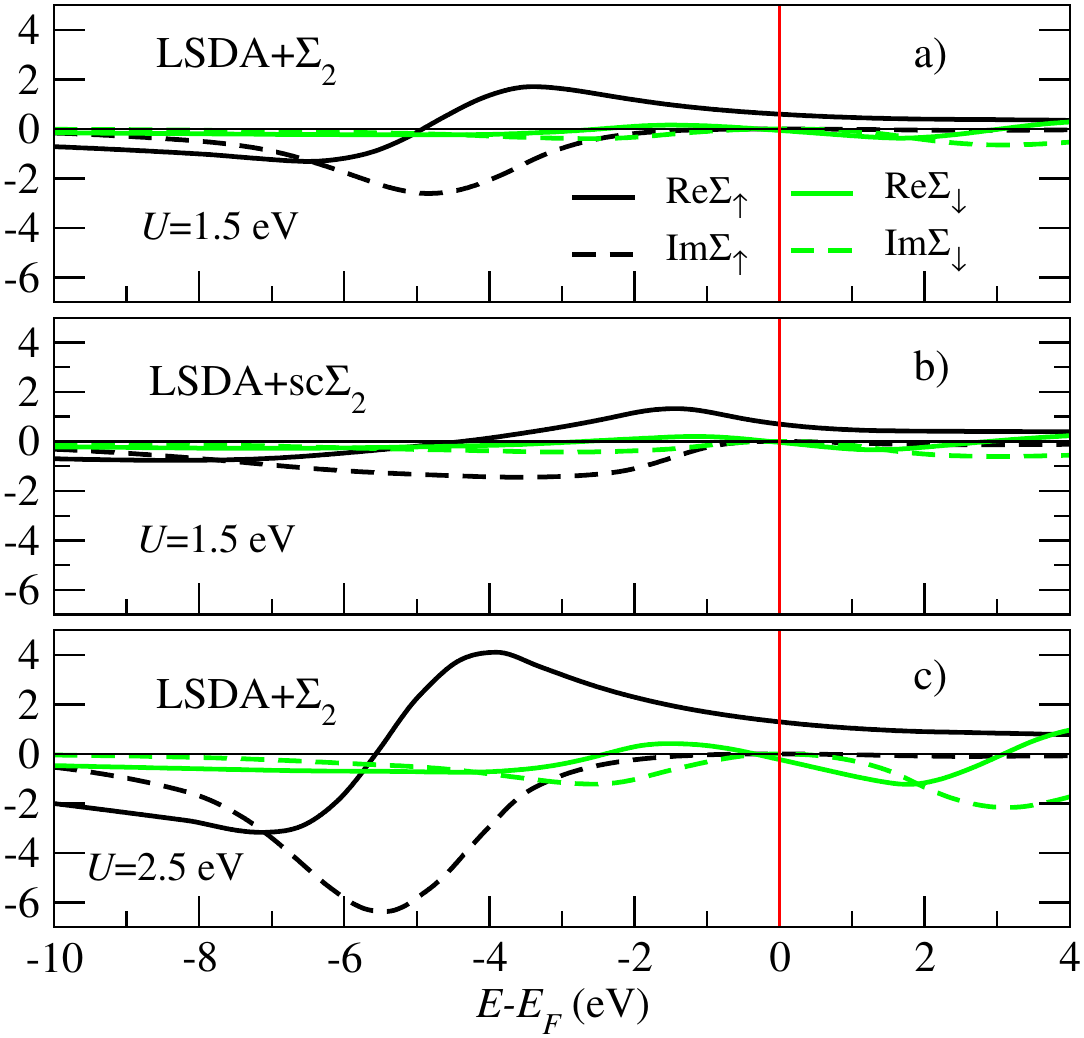}
\caption{Real and imaginary part of the self-energy (in eV) averaged over the orbital index. }
\label{fig.Sigma}
\end{figure}

The changes in the DOS induced by the dynamical correlation effects can be understood by inspecting the self-energy, which is presented in Fig.~\ref{fig.Sigma} for different cases. 
Since the crystal-field splitting is rather small, the self-energy is very similar for all orbitals. Thus, we present in Fig.~\ref{fig.Sigma} the self-energy averaged over the orbital indexes. 
The overall shape of the real and imaginary parts is typical of ferromagnetic transition metals~\cite{gr.ma.07}.
For example, in Fig.~\ref{fig.Sigma}-a we note that, 
in the spin up channel, the real part of the self-energy is positive in the energy range between $-4.5$ eV and the Fermi level, and it shows a maximum at about $E-E_F \approx-2.8$ eV. 
This causes the shift of occupied $d$ states
towards to the Fermi level as observed in Fig.~\ref{fig.DOS}. In contrast, for energies below $-4.5$ eV, the negative real part of the self-energy draws the spectral weight towards lower energies
leading to the formation of the satellite in the DOS. The imaginary part has an extended negative peak centered at $E-E_F \approx-4.5$~eV resulting in the large broadening of the satellite. Increasing the $U$ value to $2.5$ eV (Fig.~\ref{fig.Sigma}-c),  
the peak in the real part of the self-energy for spin up is enhanced to compensate the larger LSDA+U potential, 
which brings the spin up $d$ states towards too low energies. The maximum of the imaginary part also becomes more pronounced than for $U=1.5$ eV. 
In the spin down channel, correlation effects are much less pronounced for both $U=1.5$ and $U=2.5$ eV and the self-energy is quite small. \\
It is interesting to observe how the spin up self-energy changes when evaluated self-consistently (Fig.~\ref{fig.Sigma}-b). The positive maximum of the real part shifts of almost 2 eV towards the Fermi energy compared to the non-self-consistent LSDA+$\Sigma_2$ calculation. Furthermore, in the energy region below $E-E_F<-6$ eV, where the real part of the self-energy is negative, its magnitude is reduced. Similarly, the imaginary part also becomes smaller. These two factors explain why the satellite is partly suppressed in self-consistent self-energy calculations.  \\
The self-energy near the Fermi level has Fermi-liquid character: for the imaginary part, we have $-\mathrm{Im}\Sigma_\sigma(E)\propto (E-E_F)^2$, whereas the real part has negative slope, 
$\partial /\partial E (\mathrm{Re}\Sigma_{\sigma}(E))<0$.
We then evaluate the mass enhancement, which amounts to
\begin{equation}
 \bigg( \frac{m^*}{m}\bigg)_{\lambda, \sigma} =1-\frac{\partial }{\partial E } \left( \mathrm{Re}\Sigma_{\lambda,\sigma}(E) \right)
\end{equation}
where $m$ represents the band-mass obtained within the LSDA(+U) calculations. 
The enhancement factors with respect to LSDA are very similar for all $d$ orbitals and spin channels. For $U=1.5$ eV, they are on average about $1.2$, which indicates that the system is medium-correlated. When the self-energy is calculated self-consistently, $m^*/m$ becomes larger and is equal to about $1.4$. This is a due of the shift of the main peak of $\mathrm{Re}\Sigma_{\lambda,\sigma}(E)$ towards the Fermi level. Therefore we find that, while the self-consistent iterations partly suppresses many-body correlation effects at high energy, they make electrons at the Fermi energy slightly more correlated. 

\begin{figure}[h]
\centering\includegraphics[width=0.5\linewidth,clip=true]{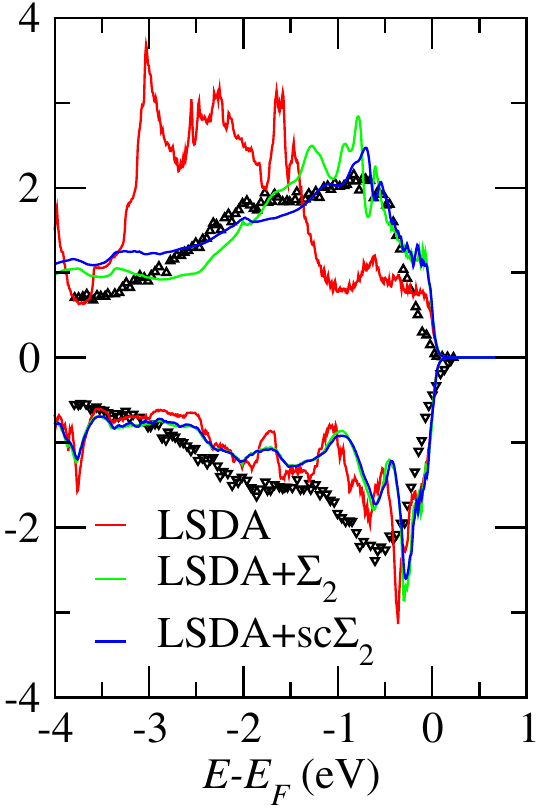}
\caption{Comparison between the calculated DOS and the photoemission spectrum of Ref. \cite{ge.ba.97} (the experimental data points are represented as small black triangles). The calculations are for $U=1.5$ eV.}
\label{fig.experiment}
\end{figure}

\subsection{Comparison to experimental data}
We now assess the performances of the various methods against experiments. Photoemission spectra of a Fe monolayer on W(110) were measured in Ref. \cite{ge.ba.97} with the electron excitation occurring along the $\Gamma$-$\Sigma$-$N$ direction of the Brillouin zone. The results are presented as small triangles in Fig. \ref{fig.experiment} along with the DOS calculated by using LSDA, LSDA+$\Sigma_2$, LSDA+sc$\Sigma2$ for $U=1.5$ eV and $J=0.5$ eV. Since electrons from both Fe and W contribute to the experimental signal, we present the sum of the DOS of the Fe monolayer and of the first W layer underneath. The DOS is convoluted with a Fermi function to introduce a smooth cut-off around the Fermi energy. We assume that the DOS can be directly compared to experiments, which means that transition matrix element effects are neglected.\\ 
The most striking observation is that LSDA drastically overestimates the spin-splitting between majority and minority states. The experimental spin up spectrum quickly rises below the Fermi energy and the maximum is at $E-E_F\approx-0.5$ eV, whereas the LSDA DOS for the spin up channel is centered at about $-2.2$ eV. The inability of LSDA to accurately predict the spin-splitting of states in metallic ferromagnetic systems has been pointed out in a number of works \cite{mo.ma.02, br.mi.06, gr.ma.07}. \\
Calculations including dynamical self-energy contributions shift the majority DOS towards the Fermi level thus correcting for the LSDA short-comes. The LSDA+$\Sigma_2$ DOS is in quantitative agreement with the experimental data. The method captures the main correlation effect in the system.\\ 
We can not identify which scheme, LSDA+$\Sigma_2$ or LSDA+sc$\Sigma_2$, performs better. Both methods give a very similar DOS in the energy range, where the experimental data is available, and the broadening of the photoemission spectrum does not allow to distinguish between small differences in the spectral distribution. Photoemission experiments probing the the energy region $E-E_F<-4$ eV and the existence of the satellite features would provide stronger indications about the accuracy of LSDA+$\Sigma_2$ and LSDA+sc$\Sigma_2$. We hope that these experiments will be performed in the future.     \\

\section{Conclusions}

DFT+$\Sigma_2$ combines DFT to a multi-orbital solver for the Hubbard model, where the Coulomb interaction parameter $U$ is treated in perturbation theory up to second order. The method provides a realistic description of moderately correlated materials. Our implementation is 
designed to investigate TM surfaces, interfaces, and layered systems in general. 
As an example of application, we considered a Fe monolayer on a W(110) substrate. We discussed in detail the correlated DOS comparing the results with those of LSDA and LSDA+U calculations. 
We found that second order self-energy contributions led to a shift of the majority-spin DOS towards the Fermi level and therefore to a reduction of the $d$ states spin-splitting compared to LSDA calculations. This result is in quantitative agreement with available photoemission spectra.\\
The use of different unperturbed states and the inclusion of the first-order contribution in DFT+$\Sigma_2$ gave differences in the DOS spin-splitting and spectral width. 
These differences were nonetheless rather minor. 
The self-consistent evaluation of the self-energy
leads to a redistribution of the spin up spectral weight,
in particular at energies far from the Fermi level. The spin-splitting of the $3d$ states changes very little during the self-consistent cycle. In contrast, a satellite feature far in energy below the Fermi level is considerably reduced owing to multi-band screening effects. \\
In our calculations we employed the local approximation thus we neglected spatially non-local correlation effects. However, the calculation of second order self-energy contribution can in principle be extended to allow for momentum dependence, although this would increase significantly the complexity of the numerical implementation and the computational overhead.\\ 
Overall, our implementation of DFT+$\Sigma_2$ can be readily used to simulate, at a relatively low computational cost, correlation effects in the electronic structure of heterostructures comprising TMs. The calculation of the self-energy on the real energy axis is particularly convenient if one is interested in treating charge and spin transport properties using generalized Landauer methods~\cite{dr.ru.17,Ja.15}, which require the evaluation of the energy dependent transmission function~\cite{ch.mo.15,book1}. As such, we believe that DFT+$\Sigma_2$ will provide important theoretical inputs for the design of materials and devices for electronics and spintronics. 

\section{Acknowledgements}
AD and AH acknowledge funding by the Science Foundation Ireland (SFI) and the Royal Society through the University Research Fellowship URF-R1-191769. 
MMR acknowledges funding provided by the Institute of Physics
Belgrade, through the grant by the Ministry of Education,
Science, and Technological Development of the Republic of
Serbia. 
IR acknowledges the support of the UK government department for Business, Energy and Industrial Strategy through the UK national quantum technologies
programme.
LC acknowledges the financial support by the Deutsche Forschungsgemeinschaft through TRR80 (project F6) Project number 107745057. 
Computational resources were provided by Trinity College Dublin Research IT. 

\appendix
\section{Second order self-energy}\label{app.diagrams}

We provide here details about the perturbative treatment of the multi-orbital electronic system. We use the Keldysh Green's functions formalism \cite{bookStefanucci}. The self-energy, up to second order can be schematically expressed using skeleton Feynman diagrams
\begin{eqnarray}
\Sigma=\Diagram{ fdV & \!\text{\CIRCLE}\! g & \!\text{\CIRCLE}\!\! & \quad\; & oAwV\\ 
fuA
}
\qquad+\quad\Diagram{ & \text{}\\ \nonumber
\,\, fdV\! & \text{\CIRCLE} & \quad\;\;\! y & \text{\CIRCLE}\\
 &  & \!\! f\!\!\!\!\!\!\! fV\!\!\!\!\!\! f\!\!\\
 &  &  &  & \!\! fdV
}
+\\ 
\quad\Diagram{ & \text{\!\!\!\!\CIRCLE} & \quad\;\; lAnV & \!\!\text{\CIRCLE}\\
 & \!\!\! gv\, &  & gv\\
fdV\quad\!\! & \!\!\! gv\text{\!\!\!\CIRCLE} & \!\!\!\! f\!\!\!\!\!\! fV\!\!\!\!\! f & gv\!\!\!\text{\CIRCLE}\\
 &  &  & fdV
}
+\quad\Diagram{ & \text{}\\
\,\, fdV\! & \text{\CIRCLE} & \quad\;\;\! y & \text{\!\!\!\!\!\CIRCLE\;\;\;\;\,\CIRCLE} & \!\!\!\! y & \!\text{\CIRCLE}\\
\\
 &  & fV & \!\! fV & \!\!\!\! fV\\
 &  &  &  &  &  & \!\! fdV
}
\ ,
\label{Eq:FeyD}
\end{eqnarray}
where the dots represents space-(contour time) coordinates, the straight line represents the dressed Green's function and the wiggly lines represents the interaction. Converted into equations these diagrams give  four contributions to the self-energy
\begin{eqnarray}
&&\Sigma_{\lambda\sigma}^{(1, 1)}({\bf k};t,t') = \nonumber \\
&&-\mathrm{i}\delta_{C}(t,t') \underset{\lambda_{f},\sigma_{f}}{\sum}\underset{{\bf p}}{\sum}\tilde{U}_{\lambda\lambda_{f}\lambda\lambda_{f}}^{\sigma\sigma_{f}\sigma\sigma_{f}}(0;t)G_{\lambda_{f}\sigma_{f}}({\bf p};t,t), \\ \nonumber \\
&&\Sigma_{\lambda\sigma}^{(1, 2)}({\bf k};t,t') = \nonumber \\ &&\mathrm{i}\delta_{C}(t,t')\underset{\lambda_{f}}{\sum}\underset{{\bf p}}{\sum}\tilde{U}_{\lambda\lambda_{f}\lambda_{f}\lambda}^{\sigma\sigma\sigma\sigma}({\bf k}-{\bf p};t)G_{\lambda_{f}\sigma}({\bf p};t,t'), \\ \nonumber \\
&&\Sigma_{\lambda_{a}\lambda_{d}\sigma_{a}\sigma_{d}}^{(2, 1)}({\bf k};t,t') = \nonumber \\ &&\underset{\lambda_b,\lambda_e,\lambda_f, \sigma_f}{\sum}\underset{{\bf p},{\bf q}}{\sum} \tilde{U}_{\lambda_{a}\lambda_{e}\lambda_{b}\lambda_{f}}^{\sigma_{a}\sigma_{f}\sigma_{a}\sigma_{f}}({\bf q};t)\tilde{U}_{\lambda_{f}\lambda_{b}\lambda_{e}\lambda_{d}}^{\sigma_{f}\sigma_{a}\sigma_{f}\sigma_{a}}({\bf q};t')\delta_{\sigma_{a}\sigma_{d}} \nonumber \\
&&G_{\lambda_b \sigma_a} ({\bf k}+{\bf q};t,t')G_{\lambda_f \sigma_f}({\bf p};t',t)G_{\lambda_e \sigma_f}({\bf p}-{\bf q};t,t'), \\ \nonumber \\
&&\Sigma_{\lambda_{a}\lambda_{d}\sigma_{a}\sigma_{d}}^{(2, 2)}({\bf k};t,t') = \nonumber \\ &&-\underset{\lambda_b,\lambda_c,\lambda_e}{\sum}\underset{{\bf p},{\bf q}}{\sum} \tilde{U}_{\lambda_{a}\lambda_{e}\lambda_{c}\lambda_{b}}^{\sigma_{a}\sigma_{a}\sigma_{a}\sigma_{a}}({\bf p};t)\tilde{U}_{\lambda_{c}\lambda_{b}\lambda_{e}\lambda_{d}}^{\sigma_{a}\sigma_{a}\sigma_{a}\sigma_{a}}({\bf q};t') \delta_{\sigma_{a}\sigma_{d}} \nonumber \\
&&G_{\lambda_b \sigma_a}({\bf k}-{\bf q};t,t')G_{\lambda_e \sigma_a}({\bf k-}{\bf p}-{\bf q};t',t)G_{\lambda_c \sigma_a}({\bf k}-{\bf p};t,t'). \nonumber \\
\end{eqnarray}
In the specific case of the Hubbard model, the interaction matrix $\tilde{U}$ is time and momentum independent and we refer to it as $U$ in the main text. Additionally, the Hubbard $U$ matrix is also spin independent, but, in this appendix, we keep the spin indices in accordance with the most general notation.\\
After applying the Langreth rules \cite{bookStefanucci} and performing the Fourier transform we obtain the first order terms
\begin{eqnarray}
\label{eq_A:HF1}
&&\Sigma_{\lambda\sigma}^{(1, 1)r}({\bf k};E)=-\mathrm{i}\underset{\lambda_{f},\sigma_{f}}{\sum}U_{\lambda\lambda_{f}\lambda\lambda_{f}}^{\sigma\sigma_{f}\sigma\sigma_{f}}\underset{{\bf p}}{\sum}\int\frac{d\epsilon_{1}}{2\pi}G_{\lambda_{f}\sigma_{f}}^{<}({\bf p};\epsilon_{1}), \nonumber \\ 
\label{eq_A:HF2}
&&\Sigma_{\lambda\sigma}^{(1, 2)r}({\bf k};E)=\mathrm{i}\underset{\lambda_{f}}{\sum}U_{\lambda\lambda_{f}\lambda_{f}\lambda}^{\sigma\sigma\sigma\sigma}\underset{{\bf p}}{\sum}\int\frac{d\epsilon_{1}}{2\pi}G_{\lambda_{f}\sigma}^{<}({\bf p};\epsilon_{1}), \nonumber \\
\end{eqnarray}
where the lesser Green's function in equilibrium assumes the following form
\begin{equation}
G^<_{\lambda \sigma}(E) = -2\mathrm{i} f(E) \; \mathrm{Im}\, G^r_{\lambda \sigma}(E),
\end{equation}
with $g^r_{\lambda \sigma}(E)$ the retarded Green's function and $f(E)$ the Fermi function. The first order terms 
can then be written in well known Hartree-Fock expression 
of Eq. (\ref{eq.first}).\\
The two second order terms giving reads
%
\begin{eqnarray}
&\Sigma_{\lambda_{a}\lambda_{d}\sigma\sigma}^{(2, 1)r}({\bf k};E)=\nonumber \\
&\underset{\lambda_{b},\lambda_{e},\lambda_{f},\sigma_{f}}{\sum}U_{\lambda_{a}\lambda_{e}\lambda_{b}\lambda_{f}}^{\sigma\sigma_{f}\sigma\sigma_{f}}U_{\lambda_{f}\lambda_{b}\lambda_{e}\lambda_{d}}^{\sigma_{f}\sigma\sigma_{f}\sigma}\underset{{\bf p},{\bf q}}{\sum}\int_{-\infty}^{\infty}\frac{d\epsilon_{1}}{2\pi}\int_{-\infty}^{\infty}\frac{d\epsilon_{2}}{2\pi}[ \nonumber \\
& G_{\lambda_{b}\sigma}^{<}({\bf k}+{\bf q};E+\epsilon_{1})G_{\lambda_{f}\sigma_{f}}^{<}({\bf p};\epsilon_{2})G_{\lambda_{e}\sigma_{f}}^{r}({\bf p}-{\bf q};\epsilon_{2}-\epsilon_{1}) \nonumber \\
& +G_{\lambda_{b}\sigma}^{r}({\bf k}+{\bf q};E+\epsilon_{1})G_{\lambda_{f}\sigma_{f}}^{<}({\bf p};\epsilon_{2})G_{\lambda_{e}\sigma_{f}}^{<}({\bf p}-{\bf q};\epsilon_{2}-\epsilon_{1}) \nonumber \\
& +G_{\lambda_{b}\sigma}^{r}({\bf k}+{\bf q};\epsilon+\epsilon_{1})G_{\lambda_{f}\sigma_{f}}^{<}({\bf p};\epsilon_{2})G_{\lambda_{e}\sigma_{f}}^{r}({\bf p}-{\bf q};\epsilon_{2}-\epsilon_{1}) \nonumber \\
& +G_{\lambda_{b}\sigma}^{<}({\bf k}+{\bf q};E+\epsilon_{1})G_{\lambda_{f}\sigma_{f}}^{A}({\bf p};\epsilon_{2})G_{\lambda_{e}\sigma_{f}}^{<}({\bf p}-{\bf q};\epsilon_{2}-\epsilon_{1})] \nonumber \\ \label{eq:sigma2_1}
\end{eqnarray}

\begin{eqnarray}
&\Sigma_{\lambda_{a}\lambda_{d}\sigma\sigma}^{(2, 2)r}({\bf k};E)= \nonumber\\ &-\underset{\lambda_{b},\lambda_{c},\lambda_{e}}{\sum}U_{\lambda_{a}\lambda_{e}\lambda_{c}\lambda_{b}}^{\sigma\sigma\sigma\sigma}U_{\lambda_{c}\lambda_{b}\lambda_{e}\lambda_{d}}^{\sigma\sigma\sigma\sigma}\underset{{\bf p},{\bf q}}{\sum}\int_{-\infty}^{\infty}\frac{d\epsilon_{1}}{2\pi}\int_{-\infty}^{\infty}\frac{d\epsilon_{2}}{2\pi}[ \nonumber \\
& G_{\lambda_{b}\sigma}^{<}({\bf k}+{\bf q};E-\epsilon_{2})G_{\lambda_{e}\sigma}^{<}({\bf p};\epsilon_{2})G_{\lambda_{c}\sigma}^{r}({\bf p}-{\bf q};\epsilon_{2}-\epsilon_{1}) \nonumber \\
& +G_{\lambda_{b}\sigma}^{r}({\bf k}+{\bf q};E-\epsilon_{2})G_{\lambda_{e}\sigma}^{<}({\bf p};\epsilon_{2})G_{\lambda_{c}\sigma}^{<}({\bf p}-{\bf q};\epsilon_{2}-\epsilon_{1}) \nonumber \\
& +g_{\lambda_{b}\sigma}^{r}({\bf k}+{\bf q};E-\epsilon_{2})G_{\lambda_{e}\sigma}^{<}({\bf p};\epsilon_{2})G_{\lambda_{c}\sigma}^{r}({\bf p}-{\bf q};\epsilon_{2}-\epsilon_{1}) \nonumber \\
& +G_{\lambda_{b}\sigma}^{<}({\bf k}+{\bf q};E-\epsilon_{2})G_{\lambda_{e}\sigma}^{A}({\bf p};\epsilon_{2})G_{\lambda_{c}\sigma}^{<}({\bf p}-{\bf q};\epsilon_{2}-\epsilon_{1})]. \nonumber \\\label{eq:sigma2_2}
\end{eqnarray}
%
After a few steps of algebra and neglecting the momentum dependence, we easily obtain Eq. (\ref{eq:sig2r}). We note that we drop the superscript ``$r$'' for ``retarded'' in the main text to keep the notation lighter. 

\section{Non-magnetic calculations}\label{app.nonmagnetic}
\begin{figure}[h]
\centering\includegraphics[width=\columnwidth,clip=true]{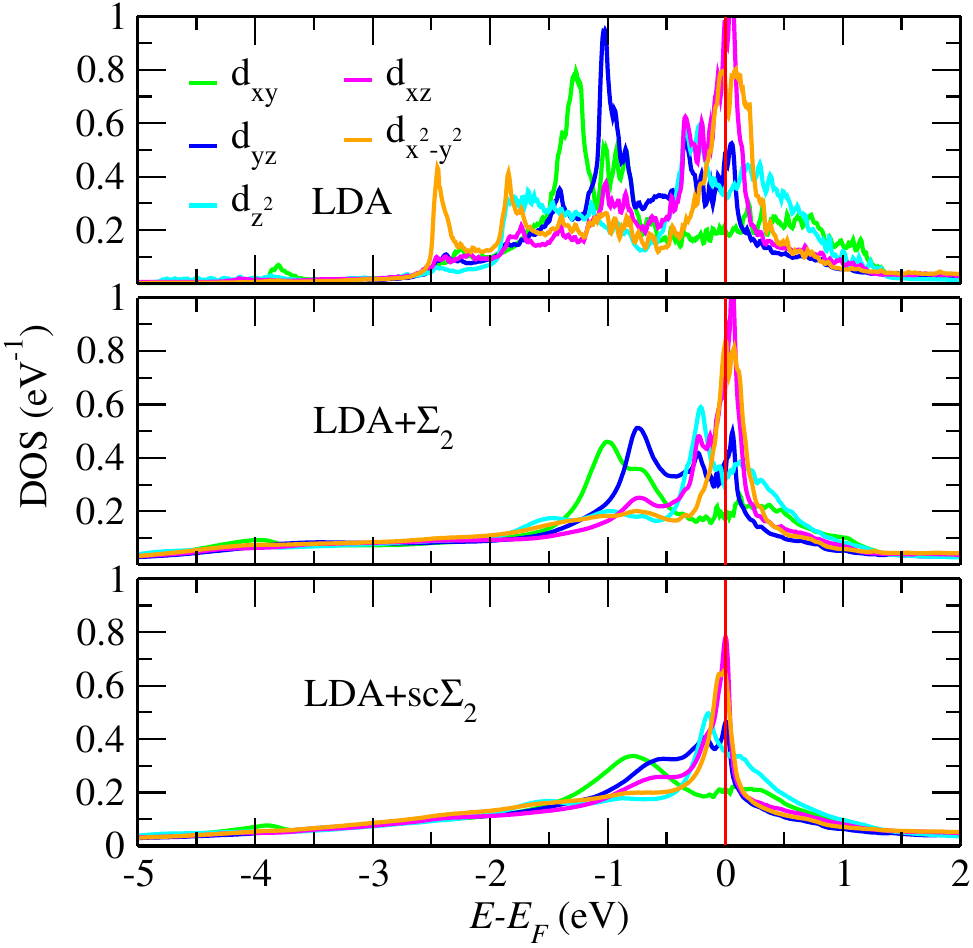}
\caption{Non-spin-polarized orbital resolved DOS of the Fe atom obtained via LDA, LDA+$\Sigma_2$ and LDA+$\Sigma_2$ calculations. The Fermi level is at 0 eV. Self-energy calculations are for $U=1.5$ eV and $J=0.5$ eV.}
\label{fig.d-resolved-DOS_para}
\end{figure}

Self-energy calculations can be carried out also for the non-magnetic phase and can be combined with DFT within the local density approximation (LDA). \\
The non-magnetic orbital-resolved DOS obtained with LDA and LDA+$\Sigma_2$ is presented in Fig. \ref{fig.d-resolved-DOS_para}. 
The general features are similar to those discussed in Sec. \ref{sec.DOS}. In LDA calculations, the DOS extends across the Fermi energy from about $-3$ eV to $1.5$ eV, i.e. over the $d$ band width equal to about $4.5$ eV. There are two peaks well below the Fermi energy, at $E-E_F\approx -1.3$ eV and $\approx -1$ eV. They stem respectively from the $d_{xy}$ and $d_{yz}$ orbitals. Instead, the $d_{xz}$ and $d_{x^2-y^2}$ orbitals give the dominant contribution to quite sharp peaks at the Fermi energy. The DOS projected over $d_{z^2}$ is much smoother. Overall the total DOS at the Fermi energy is $\mathrm{DOS}(E_F)=2.9$ eV$^{-1}$. Using the Stoner parameter $I\approx 1$ eV estimated from the band-splitting in Fig. \ref{sec.DOS} as explained in Refs. \cite{gu.78,zeller}, the Stoner criterion $I\mathrm{DOS}(E_F)>1$ is satisfied and therefore the ferromagnetic state is favored over the non-magnetic one. The same conclusion is also found by comparing the DFT total energies of the two states. 
The orbital $d_{yz}$ has the largest occupation equal to about $1.47$ electrons, while $d_{z^2}$ and $d_{x^2-y^2}$ have the lowest occupation, respectively $1.27$ and $1.25$ electrons. The orbitals $d_{xy}$ and $d_{xz}$ orbital have similar occupations of about $1.32$ electrons. \\
The second order self-energy induces some redistribution of the spectral weight. The main effect in LDA+$\Sigma_2$ is that the peaks associated to the $d_{xz}$ and $d_{x^2-y^2}$ orbitals in the DOS at the Fermi level become sharper, while the other peaks stemming from $d_{xy}$ and $d_{yz}$ are shifted towards $E_F$ by about $0.3$ eV. Furthermore, the satellite at $E-E_F<-3$ eV appears.\\
In self-consistent self-energy calculations, some further spectral redistribution occurs. The peak in the $d_{yz}$-projected DOS (blue line) at about $E-E_F\approx -0.7$ eV gets reshaped into a shoulder-like feature accompanying the main peak at $E_F$. At the same time, the peak in the $d_{xy}$-projected DOS (green line) becomes broader. As, found for the spin-polarized calculation, the satellite becomes much less marked. 

\bibliography{ref}

\end{document}